\documentclass{elsart}
\usepackage{epsfig,amsmath,amssymb,natbib}
\journal{Journal of Computational Physics}

\newcommand*{\eg}{e.g.,\ }
\newcommand*{\ie}{i.e.,\ }

\begin{document}
\bibliographystyle{elsart-harv}
\begin{frontmatter}

\title{A 3D Spectral Anelastic Hydrodynamic Code for Shearing, Stratified Flows}
\author{Joseph A. Barranco\thanksref{NSFAAPF}\thanksref{CfA}}
\address{Kavli Institute for Theoretical Physics, University of California,\\
Santa Barbara, CA 93106}
\ead{jbarranco@cfa.harvard.edu}
\thanks[NSFAAPF]{NSF Astronomy \& Astrophysics Postdoctoral Fellow}
\thanks[CfA]{Current affiliation: Harvard-Smithsonian Center for Astrophysics, 60 Garden St., MS-51, Cambridge, MA 02138}
\author{Philip S. Marcus}
\address{Dept. of Mechanical Engineering, University of California,\\
Berkeley, CA 94720}
\ead{pmarcus@me.berkeley.edu}

\begin{abstract}
We have developed a three-dimensional (3D) spectral hydrodynamic code to
study vortex dynamics in rotating, shearing, stratified systems (\eg the
atmosphere of gas giant planets,  protoplanetary disks around newly forming
protostars).   The time-independent background state is stably stratified in
the vertical direction and has a unidirectional linear shear flow aligned
with one horizontal axis.  Superposed on this background state is an
unsteady, subsonic flow that is evolved with the Euler equations subject to
the anelastic approximation to filter acoustic phenomena.  A Fourier-Fourier
basis in a set of quasi-Lagrangian coordinates that advect with the
background shear is used for spectral expansions in the two horizontal
directions.  For the vertical direction, two different sets of basis
functions have been implemented: (1) Chebyshev polynomials on a truncated,
finite domain, and (2) rational Chebyshev functions on an infinite domain.
Use of this latter set is equivalent to transforming the infinite domain to
a finite one with a cotangent mapping, and using cosine and sine expansions
in the mapped coordinate.  The nonlinear advection terms are time integrated
explicitly, whereas the Coriolis force, buoyancy terms, and pressure/enthalpy
gradient are integrated semi-implicitly.  We show that internal gravity waves
can be damped by adding new terms to the Euler equations.  The code exhibits
excellent parallel performance with the Message Passing Interface (MPI).  As 
a demonstration of the code, we simulate the merger of two 3D vortices in the
midplane of a protoplanetary disk.
\end{abstract}
\end{frontmatter}

\section{Introduction}

Three of Jupiter's notable features are: rapid rotation (spin period of just
under ten hours); an atmosphere striped with a large number of alternating
zones and belts corresponding to strongly shearing east-west winds (hundreds
of m/s); and many long-lived, coherent vortices, the most prominent being the
Great Red Spot (GRS).  Of course, these three characteristics -- rotation,
shear, and vortices -- are all dynamically linked, so much so that it is
often claimed that the presence of the first two implies the likely existence
of the third \citep[see][for reviews]{marcus90a,marcus93}.  Protoplanetary
disks (the disks of gas and dust in orbit around newly-forming protostars)
also have rapid rotation and intense shear, which has inspired proposals that
such disks should also be populated with long-lived, coherent storms
\citep{barge95,aw95,barranco00a,barranco00b,barranco05a}.  These vortices may
play two critical roles in star and planet formation: (1) In cool,
non-magnetized disks, vortices may transport angular momentum radially
outward so that mass can continue to accrete onto the growing protostar, and 
(2) vortices are very efficient at capturing and concentrating dust
particles, which may help in the formation of planetesimals, the basic
``building blocks'' of planets.

Motivated by these geophysical and astrophysical problems, we have developed 
a three-dimensional (3D) spectral hydrodynamic code that employs specially
tailored algorithms to handle the computational challenges due to rapid
rotation, intense shear, and strong stratification.  In subsonic flow,
short-wavelength acoustic waves have periods that are much shorter than the
characteristic timescale of the large-scale advective motions.  In numerical 
simulations, the time-step for an explicit algorithm must be short enough to 
temporally resolve these fast waves (\ie the Courant-Friedrich-Lewy, or CFL, 
condition), which may be inefficient for calculating the evolution of the
slower, large-scale flow for long integration times.  One strategy is to
filter sound waves from the fluid equations (``sound-proofing'') so that the 
time-step will be limited by the longer advective timescale.  The anelastic
approximation does this by replacing the full continuity equation with the
kinematic constraint that the mass flux be divergence-free.  This
approximation still allows for the effects of density stratification (\eg
buoyancy in the vertical momentum equation, pressure-volume work in the
energy/entropy equation) and has been employed extensively in the study of
deep, subsonic convection in planetary atmospheres \citep{ogura62,gough69,
bannon96b} and stars \citep{gilman81, miesch00}.  In \citet{barranco00a}, we 
re-derived the anelastic approximation in the context of protoplanetary
disks.  Stratified media support the propagation of internal gravity waves.
As these waves travel from high density to low density regions, their
amplitudes can grow to large values (so as to conserve energy flux).  If the
density contrast is large, velocity and thermodynamic fluctuations can become
sufficiently large so as to invalidate the anelastic approximation and/or
violate the CFL condition.  We have developed a technique based on ``negative
feedback'' to damp these waves when they propagate into low-density gas which
has very little inertia.

We compute the evolution of the anelastic equations with a spectral method.
The basic philosophy of spectral methods is to approximate any function of
interest with a finite sum of basis functions multiplied by spectral
coefficients \citep{gottlieborszag77,marcus86a,canuto88,boyd00}.  A partial
differential equation (PDE) in space and time is reduced to a coupled set of
ordinary differential equations (ODE) for the time evolution of the spectral
coefficients.  The chief advantage of spectral methods over finite-difference
methods is accuracy per degrees of freedom (\eg number of spectral modes
or number of grid points).  In one dimension, the global error (\eg $L_2$
norm) for a spectral method with $N$ spectral modes scales as
$(1/N)^N$, whereas for a finite-difference method with $N$ grid points, the
error scales as $(1/N)^p$, where $p$ is the (fixed) order of the method.
Thus, to get the same level of accuracy, spectral methods generally require
far fewer degrees of freedom.  This advantage is even more pronounced in 3D
problems requiring high resolution.  

Because of the linear background shear, the fluid equations are not
autonomous in the cross-stream coordinate, making it problematic to apply
periodic boundary conditions in this direction.  The equations can be made
autonomous in the horizontal directions by transforming to a set of
Lagrangian shearing coordinates \citep{goldreich65a,marcus77,rogallo81}. 
Features in the flow that are advected by the shear appear quasi-stationary
in the shearing coordinates, allowing for larger time steps to be taken in
the numerical integration.  Because the background state generally depends on
the vertical coordinate, we do not impose periodicity in the vertical
direction.  We have implemented the code with two different sets of vertical 
basis functions: (1) Chebyshev polynomials on a truncated, finite domain, and
(2) rational Chebyshev functions on an infinite domain.  Use of this latter
set is equivalent to transforming the infinite domain to a finite one with a
cotangent mapping, and using cosine and sine expansions in the mapped
coordinate \citep{cain84,boyd00}.

The code is pseudospectral.  That is, nonlinear terms are not computed via
convolution, which is computationally slow; rather, they are computed on a
grid of collocation points, and then transformed to wavenumber space with
Fast Fourier Transforms (FFTs).  Different horizontal Fourier modes interact
only through these nonlinear terms, and are otherwise independent.  This
motivated us to divide the domain into blocks along the two horizontal
directions, but not along the vertical direction.  That is, on a parallel
machine, each processor works with only a subset of the horizontal Fourier
wavenumbers, but with all of the vertical data corresponding to those
horizontal wavenumbers.  The only time in the simulation that different
processors communicate with one another is in the computation of the FFTs
needed for the nonlinear terms.  We get excellent parallel performance with
the Message Passing Interface (MPI) on the IBM Blue Horizon and Datastar
supercomputers at the San Diego Supercomputer Center.  

In \S 2, we will review the anelastic equations with shear.
In \S 3, we will describe the spectral method in detail.
In \S 4, we will present tests of accuracy and performance, and preview
potential applications.
Finally, in \S 5, we will outline future work on these applications,
as well as propose further improvements to these simulations.

\section{Anelastic Equations with Shear}

\begin{figure}
\begin{center}
\mbox{\epsfig{file=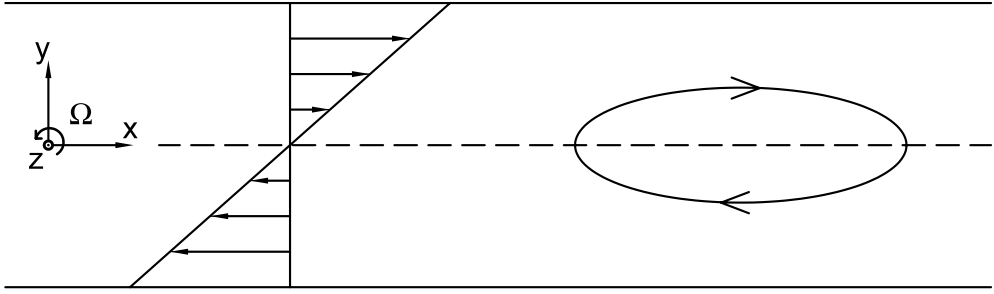,height=1.5in}}
\caption{\label{F:shearing_vortex} Linear shear flow:
$\boldsymbol{\bar{v}} = -\sigma y \boldsymbol{\hat{x}}$. 
The system rotates counterclockwise around the $z$-axis
(which points out of the page) at a rate $\Omega$.  The shear and
vortex illustrated in this example are anticyclonic: $\sigma<0$.}
\end{center}
\end{figure}

For clarity, we define up-front the following flow variables (vectors are
represented with bold type): velocity $\boldsymbol{v}$,  vorticity
$\boldsymbol{\omega}\equiv\boldsymbol{\nabla}\times\boldsymbol{v}$, pressure
$p$, density $\rho$, temperature $T$, entropy $s$, and potential temperature
$\theta$.  For an ideal gas:  $p=\rho\mathcal{R}T$ and
$s=C_v\ln(p\rho^{-\gamma})=C_p\ln\theta$,  where $C_v$ is the specific heat
at constant volume, $C_p$ is the specific heat at constant pressure,
$\mathcal{R}\equiv C_p-C_v$ is the gas constant, and $\gamma\equiv C_p/C_v$
is the ratio of specific heats.  We work in a Cartesian coordinate system
$(x,y,z)$ with corresponding unit vectors
$(\boldsymbol{\hat{x}},\boldsymbol{\hat{y}}, \boldsymbol{\hat{z}})$.  The
acceleration of gravity points in the negative $z$ direction:
$\boldsymbol{a}_{grav} = -g\boldsymbol{\hat{z}}$.  The rotation axis is
aligned with the $z$ axis: $\boldsymbol{\Omega}=\Omega\boldsymbol{\hat{z}}$.
 
Each flow variable is represented as the sum of a time-independent background
component, denoted with overbars, and a time-varying component, denoted with
tildes (\eg $\boldsymbol{v}=\boldsymbol{\bar{v}}+\boldsymbol{\tilde{v}}$).
The background velocity field is a unidirectional shear flow that is parallel
to the $x$-axis and varies only with the $y$ coordinate
(see Fig. \ref{F:shearing_vortex}):
\begin{equation}\label{E:shear_flow}
\boldsymbol{\bar{v}} = -\sigma y \boldsymbol{\hat{x}},
\end{equation}
where $\sigma$ is the constant shear rate.  The background vorticity
corresponding to this shear flow is
$\boldsymbol{\bar{\omega}} = \sigma\boldsymbol{\hat{z}}$.  The shear is
cyclonic if $\sigma$ and $\Omega$ have the same sign, and anticyclonic if
they have opposite signs. The background thermodynamic state, which we assume
depends only on the vertical coordinate $z$, is in hydrostatic balance:
\begin{equation}\label{E:hydrostatic_balance}
\frac{d\bar{p}}{dz} = -\bar{\rho}g.
\end{equation}
An important measure of the stratification is the Brunt-V\"{a}is\"{a}l\"{a}
frequency, or the frequency of buoyancy-driven oscillations in a convectively
stable atmosphere:
\begin{equation}\label{E:brunt_vaisala}
\omega_B^2(z) \equiv \frac{g}{C_p}\frac{d\bar{s}}{dz} \equiv \frac{g}{\bar{\theta}}\frac{d\bar{\theta}}{dz}.
\end{equation}

There are many different variations of the anelastic approximation
\citep{ogura62,gough69,gilman81,bannon96b}.  The key assumption they all
share is that the flow is subsonic and the thermodynamic fluctuations are
small relative to their background values:
\begin{equation}\label{E:mach_number}
\epsilon^2\equiv\left(\frac{\tilde{v}}{c_s}\right)^2\sim\frac{\tilde{p}}{\bar{p}}\sim\frac{\tilde{\rho}}{\bar{\rho}}\sim\frac{\tilde{T}}{\bar{T}}\sim\frac{\tilde{\theta}}{\bar{\theta}}\ll 1,
\end{equation}
where $\epsilon$ is the Mach number and $c_s$ is the local sound speed.  In a
typical derivation, all variables are expressed as asymptotic series
expansions in powers of the Mach number $\epsilon$, the expansions are
substituted into the governing fluid equations, and terms are grouped by like
powers of $\epsilon$.  In \citet{barranco00a}, we derived a version of the
anelastic approximation for a protoplanetary disk with a constant temperature
background.  Here, we present a more general version proposed by
\citet{bannon96b}, but modified to include a background shear:
\begin{subequations}\label{E:anelastic}
\begin{align}
0 &= \frac{1}{\bar{\rho}}\boldsymbol{\nabla}\cdot\bar{\rho}\boldsymbol{\tilde{v}} = \boldsymbol{\nabla}\cdot\boldsymbol{\tilde{v}} + \left(\frac{d\ln\bar{\rho}}{dz}\right)\tilde{v}_z, \label{E:divergence}\\
\left(\frac{\partial}{\partial t} -\sigma y\frac{\partial}{\partial x}\right)\boldsymbol{\tilde{v}} &= \sigma\tilde{v}_y\boldsymbol{\hat{x}} + \boldsymbol{\tilde{v}}\times(\boldsymbol{\tilde{\omega}}+2\Omega\boldsymbol{\hat{z}})-\boldsymbol{\nabla}\tilde{h}+\frac{\tilde{\theta}}{\bar{\theta}}g\boldsymbol{\hat{z}},\label{E:momentum}\\
\left(\frac{\partial}{\partial t} -\sigma y\frac{\partial}{\partial x}\right)\tilde{\theta}                &= -\boldsymbol{\tilde{v}}\cdot\boldsymbol{\nabla}\tilde{\theta}-\frac{d\bar{\theta}}{dz}\tilde{v}_z,\label{E:entropy}\\
\tilde{h}&=\tilde{p}/\bar{\rho}+ \tilde{v}^2/2,\label{E:enthalpy}\\
\tilde{p}/\bar{p} &= \tilde{\rho}/\bar{\rho} + 
\tilde{T}/\bar{T},\label{E:linearized_ideal_gas}\\
\tilde{\theta}/\bar{\theta} &= \tilde{p}/(\bar{\rho}gH_{\rho}) - \tilde{\rho}/\bar{\rho}, \label{E:linearized_entropy}
\end{align}
\end{subequations} 
where we have defined a dynamic enthalpy $\tilde{h}$, and a density scale
height $H_{\rho}\equiv\vert\left(d\ln\bar{\rho}/dz\right)^{-1}\vert$.  The
error in the anelastic approximation due to the neglected terms scales as
$\mathcal{O}(\epsilon^2)$.  The careful reader may note that equation
\eqref{E:linearized_entropy} is not the ``correct'' linearization of the
potential temperature;  the denominator in the pressure term should be
$\gamma\bar{p}$.  \citet{bannon96b} shows that the anelastic equations do not
conserve energy unless one makes the replacement
$\gamma\bar{p}\rightarrow\bar{\rho}gH_{\rho}$.

In the case where the background is constant temperature $\bar{T}=T_0$, we
can eliminate the potential temperature in favor of the gas temperature via
\citep{bannon96b,barranco00a}:
\begin{subequations}\label{E:constant_temperature}
\begin{align}
\frac{\tilde{\theta}}{\bar{\theta}}&\rightarrow\frac{\tilde{T}}{T_0},\\
\left(\frac{\partial}{\partial t} -\sigma y\frac{\partial}{\partial x}\right)\tilde{T} &= -\boldsymbol{\tilde{v}}\cdot\boldsymbol{\nabla}\tilde{T}-\frac{\omega_B^2}{g}\left(T_0+\tilde{T}\right)\tilde{v}_z.\label{E:temperature}
\end{align}
\end{subequations}

It is useful for diagnostic purposes to identify energy-like conserved
quantities for the anelastic system \eqref{E:anelastic}.  We define kinetic,
shear, and thermal energies:
\begin{subequations}\label{E:energy_defs}
\begin{align}
E_K &\equiv\int_V \bar{\rho}\tilde{v}^2/2 \; dV, \label{E:KE}\\
E_S &\equiv\int_V \bar{\rho}\bar{v}_x\tilde{v}_x \;dV, \label{E:SE}\\
E_T &\equiv\int_V C_p\bar{\rho}\bar{T}\left(\tilde{\theta}/\bar{\theta}\right) \; dV, \label{E:TE}
\end{align}
\end{subequations}
where we integrate over the computational volume $V$:
$-L_x/2\le x\le L_x/2$, $-L_y/2 \le y \le L_y/2$, and $-L_z \le z \le L_z$.
For this exercise, we assume that all quantities with a tilde are periodic
in $x$ and $y$.\footnote{We have not yet introduced the shearing coordinates,
but if one did this exercise assuming that variables with tildes are
periodic in the shearing coordinates, one would get the exact same result.}
We also assume that if $L_z$ is finite, then $\tilde{v}_z=0$ at $z=\pm L_z$,
or if $L_z\rightarrow\infty$, then $\tilde{v}_z\rightarrow 0$ sufficiently
fast.  To obtain an equation for the evolution of the kinetic energy, take
the dot product of the momentum flux $\bar{\rho}\boldsymbol{\tilde{v}}$ with
the momentum equation \eqref{E:momentum} and integrate over the domain $V$.
Many of the resulting terms can be written as perfect divergences which will
integrate to zero with periodicity in $x$ and $y$ and the vertical boundary
conditions.  The terms that survive the integration are source/sink terms for
the kinetic energy.  For the evolution of the shear energy, multiply the
$x$-component of the momentum equation \eqref{E:momentum}
by $\bar{\rho}\bar{v}_x$, and integrate over $V$.  Here, the integration is
not so simple because $y$ times a tilde quantity is not necessarily periodic
in $y$; this will lead to surface terms that do not cancel. For the evolution
of the thermal energy, multiply the potential temperature equation
\eqref{E:entropy} by $C_p\bar{\rho}\bar{T}/\bar{\theta}$ and integrate over
$V$.  The resulting energy evolution equations are:
\begin{subequations}\label{E:energy_evolution}
\begin{alignat}{4}
dE_K/dt &= && +\mathcal{E}_1 && +\mathcal{E}_2 && {}, \label{E:dKEdt}\\ 
dE_S/dt &= && -\mathcal{E}_1 && {} && +\mathcal{E}_3, \label{E:dSEdt} \\ 
dE_T/dt &= &&{}&&-\mathcal{E}_2&& {}, \label{E:dTEdt}\\
d(E_K + E_S + E_T)/dt &= &&{}&&{}&&+\mathcal{E}_3,\label{E:dTOTdt}
\end{alignat}
\end{subequations}
where the energy source/sink terms are defined:
\begin{subequations}\label{E:source_sinks}
\begin{align}
\mathcal{E}_1 &\equiv\int_V \sigma\bar{\rho}\tilde{v}_x\tilde{v}_y \; dV, \label{E:E1}\\
\mathcal{E}_2 &\equiv\int_V \bar{\rho}\left(\tilde{\theta}/\bar{\theta}\right)g\tilde{v}_z \; dV, \label{E:E2}\\
\mathcal{E}_3 &\equiv\int_{y=L_y/2} \sigma\bar{\rho}\tilde{v}_x\tilde{v}_y \;L_ydxdz. \label{E:E3}
\end{align}
\end{subequations}
There are two source/sink terms for the kinetic energy: exchange with the
background shear, and exchange with the thermal energy.  As expected, these
same terms appear, with opposite signs, in the evolution equations for the
shear and thermal energies.  We can define a ``total'' energy which is the
sum of kinetic, shear, and thermal energies.  The evolution equation for the
total energy has only one source/sink term, which is the surface term from
the shear energy evolution equation.  This term represents the flow of energy
into and out of the edges of the domain in the cross-stream direction.

\section{Spectral Method}

Let $q(x,y,z,t)$ represent some arbitrary variable, which may be either a
scalar or a component of a vector field.  We write:
\begin{equation}\label{E:spectral}
q(x,y,z,t) \approx \sum_{\ell=-\frac{N_x}{2}+1}^{N_x/2}\; \sum_{m=-\frac{N_y}{2}+1}^{N_y/2}\; \sum_{n=0}^{N_z}\hat{q}_{\ell m n}(t)e^{ik_x x}e^{ik_y y}\phi_n(z),
\end{equation}
where $\ell,m,n$ are integers, and $\{\hat{q}_{\ell m n}(t)\}$ are the set of
spectral coefficients.  The Fourier-Fourier basis in $x$ and $y$ has
wavenumbers:  $\{k_x \equiv \ell\Delta k_x$, $k_y \equiv m\Delta k_y$\},
where $\Delta k_x \equiv 2\pi/L_x$, $\Delta k_y \equiv 2\pi/L_y$, and where
$L_x$ and $L_y$ are the dimensions of the periodic box.  The basis functions 
for the vertical direction, $\phi_n(z)$, will be defined in Section 3.2.  If 
$q$ is a real function (and if the $\phi_n(z)$ are real functions of $z$)
then we have the reality condition:
$\hat{q}_{\ell m n}^{*} = \hat{q}_{-\ell,-m,n}$,  where $()^{*}$ represents
complex conjugation.  We only need to keep track of half the spectral
coefficients, say, for $\ell \ge 0$.

Our approach is a pseudospectral method.  Products of two or more
variables are \textit{not} computed via convolution, which is computationally
expensive.  Instead, we define a grid of collocation points:
$\{x_{\ell}\equiv\ell\Delta x,y_m\equiv m\Delta y,z_n\}$, where
$\Delta x \equiv L_x/N_x$ and $\Delta y \equiv L_y/N_y$.  The collocation
points are uniformly spaced in $x$ and $y$, but not in $z$ (see Section 3.2).
A variable can equally be represented by its set of spectral coefficients
$\{\hat{q}_{\ell m n}\}$ (``wavenumber space'' or ``FFF-space'') or by the
values of the variable at the set of grid points $\{q(x_{\ell},y_m,z_n)\}$
(``physical space'' or ``PPP-space'').  We go between these two
representations with the use of Fast Fourier Transforms (FFTs):
\begin{equation}\label{E:transform}
\{\hat{q}_{\ell m n}\} \overset{FFT}{\Longleftrightarrow} \{q(x_{\ell},y_m,z_n)\}.
\end{equation}
To compute a product of two or more variables, we transform to physical
space, multiply the variables at each collocation point, and then transform
back to wavenumber space.  The product will be contaminated with aliasing
errors, but these will be small if the variables are well-resolved (that is, 
the spectral coefficients for high wavenumbers are sufficiently small).
Finally, we mention that at times we will work in ``mixed'' space in which
the horizontal directions are in Fourier-Fourier space, but the vertical
direction is untransformed: $\{\hat{q}_{\ell m}(z_n)\}$.  We call this
``FFP-space.''

\subsection{Shearing coordinates}

\begin{figure}
\begin{center}
\mbox{\epsfig{file=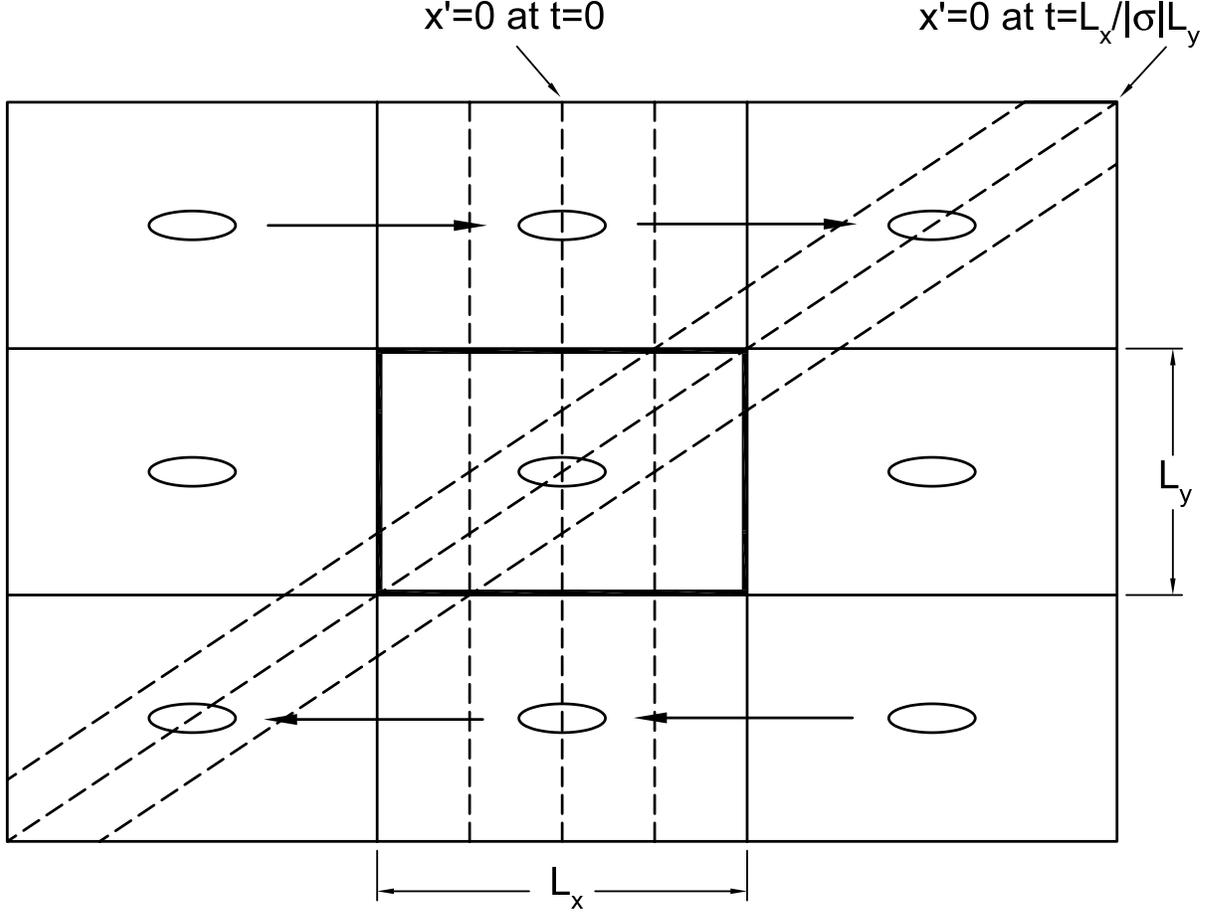,height=5in}}
\caption{\label{F:shearing_coords}
Lagrangian shearing coordinates that advect with the shear.
In this example, $\sigma<0$.  Center box is computational domain of size $(L_x,L_y)$.  Eight periodic images are also shown.  The dashed vertical lines correspond to lines of constant $x'$ at time $t=0$.  At time $t=t_{rm}\equiv L_x/|\sigma| L_y$,
the periodic images, which are carried with the shearing coordinates,
have realigned -- this is the optimal time to re-map the sheared
coordinates back onto a Cartesian grid.}
\end{center}
\end{figure}

The anelastic equations with shear \eqref{E:anelastic} are autonomous in time
$t$ and the streamwise direction $x$, but are non-autonomous in the
cross-stream direction $y$  and the vertical direction $z$.  It would be
ideal if we could make the equations autonomous in $y$ before imposing
periodicity.  Fortunately, this can be done by transforming to a set of
quasi-Lagrangian coordinates that advect with the background shear
\citep{goldreich65a,marcus77,rogallo81}:
\begin{subequations}\label{E:shearing_coords}
\begin{align}
t' &\equiv t, \\
x' &\equiv x + \sigma y t, \\
y' &\equiv y, \\
z' &\equiv z.
\end{align}
\end{subequations}
The Jacobian for this coordinate transformation is unity; the volume of a
computational cell is preserved as it is sheared.  Partial derivatives in the
two coordinate systems are related via the chain rule:
\begin{subequations}\label{E:shearing_derivs}
\begin{align}
\frac{\partial}{\partial t} &= \frac{\partial}{\partial t'} + \sigma y'\frac{\partial}{\partial x'}, \\
\frac{\partial}{\partial x} &= \frac{\partial}{\partial x'}, \\
\frac{\partial}{\partial y} &= \frac{\partial}{\partial y'} + \sigma t'\frac{\partial}{\partial x'}, \\
\frac{\partial}{\partial z} &= \frac{\partial}{\partial z'},\\
\boldsymbol{\nabla}  &= \boldsymbol{\nabla'} + \boldsymbol{\hat{y}}\sigma t'\frac{\partial}{\partial x'}.
\end{align}
\end{subequations}
The anelastic equations in the the shearing coordinates are:
\begin{subequations}\label{E:anelastic2}
\begin{align}
0 &= \left[\left(\boldsymbol{\nabla'} + \boldsymbol{\hat{y}}\sigma t'\frac{\partial}{\partial x'}\right) + \boldsymbol{\hat{z}}\left(\frac{d\ln\bar{\rho}}{dz}\right)\right]\cdot \boldsymbol{\tilde{v}}, \label{E:divergence2}\\
\frac{\partial\boldsymbol{\tilde{v}}}{\partial t'} &= \sigma\tilde{v}_y\boldsymbol{\hat{x}} + \boldsymbol{\tilde{v}}\times(\boldsymbol{\tilde{\omega}}+2\Omega\boldsymbol{\hat{z}})-\left(\boldsymbol{\nabla'}+\boldsymbol{\hat{y}}\sigma t'\frac{\partial}{\partial x'}\right)\tilde{h} + \frac{\tilde{\theta}}{\bar{\theta}}g\boldsymbol{\hat{z}},\label{E:momentum2}\\
\frac{\partial\tilde{\theta}}{\partial t'}  &= -\boldsymbol{\tilde{v}}\cdot\left(\boldsymbol{\nabla'} + \boldsymbol{\hat{y}}\sigma t'\frac{\partial}{\partial x'}\right)\tilde{\theta}-\frac{d\bar{\theta}}{dz}\tilde{v}_z,\label{E:entropy2}
\end{align}
\end{subequations}
where the vorticity is now: $\boldsymbol{\tilde{\omega}}\equiv\left(\boldsymbol{\nabla'}+\boldsymbol{\hat{y}}\sigma t'\partial/ \partial x'\right)\times\boldsymbol{\tilde{v}}$.
The equations are autonomous in both $x'$ and $y'$, at the expense that the
equations explicitly depend on time $t'$.

Variables are represented by Fourier-Fourier expansions in the shearing
coordinates, not in the original coordinates.  From the point of view of the
original coordinates, the computational grid is sheared with the background
flow (see Fig. \ref{F:shearing_coords}).  If left unchecked, the
computational grid will become greatly distorted in time: lines of constant
$x'$ will approach being parallel with lines of constant $y'$.  Periodically,
it will be necessary to re-map the shearing coordinate system back onto a
rectangular Cartesian grid \citep{rogallo81}.  This should not be done at any
arbitrary time.  In general, variables will not be periodic with respect to
the original coordinates except at those special times when the periodic
images line up again vertically.  We define this to be the re-map time:
\begin{equation}\label{E:remap1}
t_{rm} \equiv \frac{L_x}{|\sigma|L_y}.
\end{equation}

In physical space, the coordinates of a computational cell after re-mapping
$(x_{rm},y_{rm})$ can be found by inverting \eqref{E:shearing_coords} and
evaluating at the re-map time:
\begin{subequations}\label{E:remap2}
\begin{align}
x_{rm} &= x' - \sigma y' t_{rm} = x' \mp \frac{L_x}{L_y}y',\\
y_{rm} &= y',
\end{align}
\end{subequations}
where the upper sign corresponds to $\sigma>0$ and the lower sign corresponds
to $\sigma<0$.  Using the well-known Fourier shift theorem, the re-map
process can also be done in FPP-space ($x$ direction is in Fourier space, but
the $y$ and $z$ directions in physical space):  
\begin{align}\label{E:remap4}
\hat{q}_{\ell}^{rm}(y,z,t_{rm}) &= \hat{q}_{\ell}(y',z',t_{rm})\exp\left(ik_x\sigma yt_{rm}\right)\notag \\
{} &= \hat{q}_{\ell}(y',z',t_{rm})\exp\left(\pm 2\pi i\ell y/L_y\right), 
\end{align}
where the top sign corresponds to $\sigma>0$, and the bottom sign corresponds
to $\sigma<0$.

We re-map the three components of velocity and the potential temperature, but
not the dynamic enthalpy.  In an anelastic code, the enthalpy is not a
dynamic variable; rather, it adjusts instantaneously to maintain the
divergence-free condition on the momentum.  The enthalpy should be recomputed
from the Poisson-like equation after the momentum and potential temperature
have been re-mapped (see Section 3.3).

\subsection{Vertical basis functions}

\begin{figure}
\begin{center}
\mbox{\epsfig{file=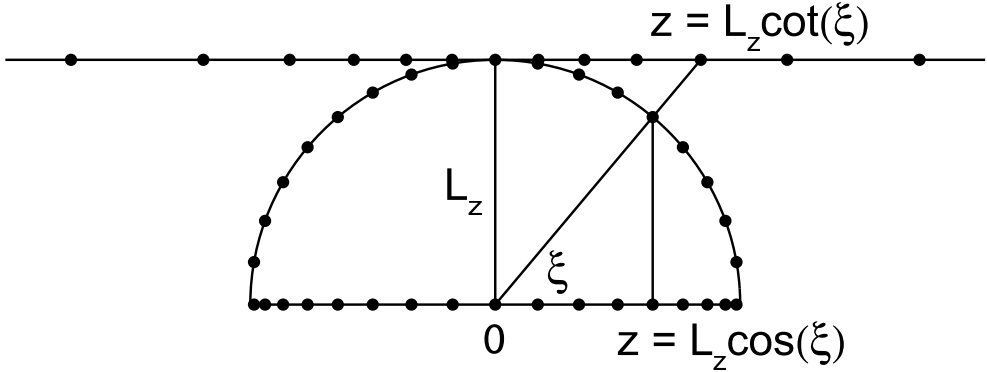,height=2in}}
\caption{\label{F:mapping}
Cosine mapping versus cotangent mapping.  The coordinate $\xi$ is
uniformly spaced from $0$ to $\pi$.  Cosine mapping clusters collocation
points around boundaries, whereas cotangent mapping concentrates collocation points around center of the domain.}
\end{center}
\end{figure}

In general, the background state may depend explicitly on the vertical
coordinate $z$.  Rather than imposing artificial periodicity in the vertical
direction, we employ two different domain mapping methods
\citep{canuto88,boyd00}.  The first mapping that we use is a cosine mapping:
\begin{equation}\label{E:mapping1}
z = L_z\cos\xi, \quad 0 \le \xi \le \pi,
\end{equation}
which truncates the vertical domain to be within $|z| \le L_z$.  When we
couple this mapping with a cosine series expansion, the resulting basis
functions are the Chebyshev polynomials:
\begin{equation}\label{E:Chebyshev_polynomials}
T_n\left(\frac{z}{L_z}\right) \equiv \cos(n\xi)\equiv\cos(n\cos^{-1}\frac{z}{L_z}).
\end{equation}
That is, use of a cosine basis in the mapped coordinate $\xi$ is completely
equivalent to using a Chebyshev basis in the original coordinate $z$.  One
disadvantage of this approach is that the collocation points (which are
uniformly spaced in the mapped coordinate $\xi$) are clustered near the
boundaries $z=\pm L_z$ (see Fig. \ref{F:mapping}).  Another disadvantage is
the need to impose what may be unphysical boundary conditions at the
boundaries $z=\pm L_z$.

The second mapping we use is a cotangent mapping \citep{cain84,boyd00}:
\begin{equation}\label{E:mapping2}
z = L_z\cot\xi, \quad 0 \le \xi \le \pi,
\end{equation}
which allows us to treat the entire infinite domain $-\infty < z < \infty$.
With this mapping, half of the collocation points are within $|z| \le L_z$,
and they are more concentrated around $z=0$ (see Fig. \ref{F:mapping}).  The
other half of the collocation points are widely spaced beyond $|z|>L_z$.  We
couple this mapping with both cosine and sine series:
\begin{subequations}\label{E:rational_Chebyshev}
\begin{align}
C_n\left(\frac{z}{L_z}\right) &\equiv \cos(n\xi)\equiv\cos(n\cot^{-1}\frac{z}{L_z}), \\
S_n\left(\frac{z}{L_z}\right) &\equiv \sin(n\xi)\equiv\sin(n\cot^{-1}\frac{z}{L_z}).
\end{align}
\end{subequations}
\citet{boyd00} calls these functions ``rational Chebyshev functions''
(although only the functions for even $n$ are rational, whereas the ones for 
odd $n$ are irrational functions). 

Why do we couple the cotangent mapping with \textit{both} cosine and sine
expansions, whereas with the cosine mapping, we use only cosine expansions?
The derivative of the $nth$ Chebyshev polynomial is (with $L_z=1$):
\begin{equation}\label{E:Cheb_deriv}
\frac{dT_n(z)}{dz} = \frac{n\sin(n\xi)}{\sin(\xi)}=2n\sum_{\substack{j=0\\ j+n\; \text{odd}}}^{n-1}\alpha_j \cos(j\xi) = 2n\sum_{\substack{j=0\\j+n\; \text{odd}}}^{n-1}\alpha_j T_j(z),
\end{equation}
where $\alpha_0=1/2$ and $\alpha_j=1$ for $j>0$.  That is, the derivative of
a Chebyshev polynomial is another Chebyshev polynomial of one lower degree.
On the other hand, the derivatives of the rational Chebyshev functions are
(with $L_z=1$):
\begin{subequations}\label{E:ratCheb_deriv}
\begin{align}
\raisetag{14pt}
\begin{split}
\frac{dC_n(z)}{dz} &= n\sin(n\xi)\sin^2(\xi)= n\{-\sin[(n\!-\!2)\xi]+2\sin(n\xi)-\sin[(n\!+\!2)\xi]\}\\ &= n\left[-S_{n-2}(z)+2S_n(z)-S_{n+2}(z)\right],
\end{split}\label{E:ratChebC_deriv}\\
\raisetag{14pt}
\begin{split}
\frac{dS_n(z)}{dz} &= -n\cos(n\xi)\sin^2(\xi)= -n\{-\cos[(n\!-\!2)\xi]+2\cos(n\xi)-\cos[(n\!+\!2)\xi]\}\\ &= -n\left[-C_{n-2}(z)+2C_n(z)-C_{n+2}(z)\right].
\end{split}\label{E:ratChebS_deriv}
\end{align}
\end{subequations}
The derivative of a rational Chebyshev function of the cosine kind is a
rational Chebyshev function of the sine kind, and vice versa.  In principle, 
one could represent the derivative of $C_n(z)$ with a convergent infinite
series of $C_n(z)$, obviating the need for $S_n(z)$.  However, we prefer to
work with recursion relations with finite numbers of terms, such as
\eqref{E:ratCheb_deriv}.  Other recursion relations for the rational
Chebyshev functions can be found in Appendix C.

\citet{boyd00} has an extensive discussion on the asymptotic
behavior of the rational Chebyshev functions.  We want to point out that
the choice of cosine or sine kind in representing a given function is 
\textit{not} determined by the parity of the function (\ie whether it is
an even or odd function), but on the asymptotic behavior of the function
at infinity.  The rational Chebyshev functions of the cosine kind approach
their asymptotic values in even powers of $1/y$, whereas those of the
sine kind approach their asymptotic values in odd powers of $1/y$.

The product of two variables represented by the same kind of rational
Chebyshev functions results in a series of the cosine kind,  whereas the
product of two variables represented by different kinds of rational Chebyshev
functions results in a series of the sine kind.  Multiplying a variable by a
function that is odd with respect to $z$ causes its series expansion to
switch kinds.  Every term in a given equation should be represented by the
same kind of expansion.  Thus, when we use the rational Chebyshev
functions for the vertical basis functions, $\tilde{v}_x$, $\tilde{v}_y$,
$\tilde{\omega}_z$, $\tilde{\theta}$, and $\tilde{h}$ are represented with
rational Chebyshev functions of the cosine kind, whereas $\tilde{v}_z$,
$\tilde{\omega}_x$, and $\tilde{\omega}_y$ are represented with rational
Chebyshev functions of the sine kind.  

\subsection{Time integration algorithms}

In this section, we describe algorithms for integrating the anelastic
equations forward in time.  Let the hat symbol (\eg $\hat{q}$) denote a
variable whose horizontal directions are in Fourier space (for clarity, we
drop the subscripts $\ell, m$ previously introduced).  One may think of
$\hat{q}$ as a either a column vector of the values of the variable at the
different vertical positions (\ie FFP-space), or a column vector of vertical 
spectral coefficients (\ie FFF-space); which of these is intended will be
clear from the context.

The equations are integrated in time via the method of ``fractional steps.''
The velocity is integrated in four steps: an advection step, a hyperviscosity
step, an explicit pressure step, and an implicit pressure step. The potential
temperature is integrated in two steps: an advection step and a
hyperdiffusion step.  We use a superscript $N$ to denote a
variable at the $N$th timestep (\eg $t^N\equiv N\Delta t$, $\boldsymbol{\hat{v}}^{N}\equiv\boldsymbol{\hat{v}}(t\!=\!t^N)$).  A fraction in the
superscript denotes a variable at an intermediate stage of the integration,
not after some fraction of a timestep (\eg $\boldsymbol{\hat{v}}^{N+1/4}$ is
the velocity after the first of four intermediate steps).

We have implemented two versions of the advection step, both based on an
explicit second-order Adams-Bashforth method.  In one method, the nonlinear
advection terms are treated explicitly and the linear terms (Coriolis and
buoyancy) are treated semi-implicitly.  In the other method, the advection,
Coriolis, and buoyancy terms are all treated explicitly, but with additional
terms that damp internal gravity waves in low-density regions.  For the
pressure step, we use the semi-implicit second-order Crank-Nicholson method. 
We have confirmed that there are no low order splitting errors, so that the
algorithms are globally second-order accurate.

\subsubsection{Step 1: Advection --- Semi-implicit treatment of
Coriolis and buoyancy forces}

The Coriolis force couples the horizontal components of the velocity,
resulting in inertial oscillations with the Coriolis frequency (modified by
linear shear): $\omega_C\equiv\sqrt{2\Omega(2\Omega+\sigma)}$.  Similarly,
buoyancy and pressure-volume work couple the vertical velocity and the
potential temperature, resulting in buoyant oscillations (\eg internal
gravity waves) with the Brunt-V\"{a}is\"{a}l\"{a} frequency $\omega_B$
\citep{kundu90,pedlosky79}.  Both inertial and buoyant oscillations can be
described by the simple set of coupled ordinary differential equations:
\begin{subequations}
\begin{align}
\dot{u}_1 &= +\alpha_1 u_2 + f_1,\\
\dot{u}_2 &= -\alpha_2 u_1 + f_2,
\end{align}
\end{subequations}
where a dot denotes time differentiation, $\alpha_1>0$, $\alpha_2>0$, and
$f_1$, $f_2$ are forcing constants.  The exact solution for arbitrary time
$t$ is:
\begin{subequations}
\begin{align}
\begin{split}
u_1(t) = & u_1(0)\cos(\omega t)
+ u_2(0)\left(\frac{\alpha_1}{\omega}\right)\sin(\omega t)\\
& + f_1t\left[\frac{\sin(\omega t)}{\omega t}\right] 
+ f_2\left(\frac{\alpha_1 t^2}{2}\right)\left[\frac{2(1-\cos(\omega t))}{(\omega t)^2}\right],
\end{split}\\
\begin{split}
u_2(t) = & u_2(0)\cos(\omega t)
- u_1(0)\left(\frac{\alpha_2}{\omega}\right)\sin(\omega t)\\
& + f_2t\left[\frac{\sin(\omega t)}{\omega t}\right] 
- f_1\left(\frac{\alpha_2 t^2}{2}\right)\left[\frac{2(1-\cos(\omega t))}{(\omega t)^2}\right],
\end{split}
\end{align}
\end{subequations}
where we have defined $\omega^2\equiv\alpha_1\alpha_2$.

We use the above solution as a template for a new integration scheme in which
the forcing terms (\eg $f_1$, $f_2$) are not constants in time, but
correspond to nonlinear advection terms, which we treat with an
Adams-Bashforth scheme:
\begin{subequations}
\begin{align}
\boldsymbol{\hat{\mathfrak{N}}} \equiv & \frac{3}{2}\left[\widehat{\left(\boldsymbol{\tilde{v}\times\tilde{\omega}}\right)}^{N}\right]
-\frac{1}{2}\left[\widehat{\left(\boldsymbol{\tilde{v}\times\tilde{\omega}}\right)}^{N-1}\right],\\
\hat{\mathfrak{M}} \equiv & \frac{3}{2}\left[-\widehat{\left(\boldsymbol{\tilde{v}\cdot\nabla}\tilde{\theta}\right)}^{N}\right]
-\frac{1}{2}\left[-\widehat{\left(\boldsymbol{\tilde{v}\cdot\nabla}\tilde{\theta}\right)}^{N-1}\right].
\end{align}
\end{subequations}
The velocity and potential temperature are updated via:
\begin{subequations}
\begin{align}
\begin{split}
\hat{v}_{x}^{N+\frac{1}{4}} = &\cos(\omega_C\Delta t)\hat{v}_{x}^{N}
+ \Delta t\,q_1(\omega_C\Delta t) \left[(2\Omega+\sigma)\hat{v}_y^N+\hat{\mathfrak{N}}_{x}\right] \\
& + \tfrac{1}{2}\Delta t^2q_2(\omega_C\Delta t)(2\Omega+\sigma)\left[
\hat{\mathfrak{N}}_{y} - i(k'_y+\sigma tk'_x)\hat{h}^{N}\right],
\end{split}\\
\begin{split}
\hat{v}_{y}^{N+\frac{1}{4}} = &\cos(\omega_C\Delta t)\hat{v}_{y}^{N}
+ \Delta t\,q_1(\omega_C\Delta t) \left[(-2\Omega)\hat{v}_x^N+\hat{\mathfrak{N}}_{y}\right] \\
& + \tfrac{1}{2}\Delta t^2q_2(\omega_C\Delta t)(-2\Omega)\left[
\hat{\mathfrak{N}}_{x} - ik'_x\hat{h}^{N}\right],
\end{split}\\
\begin{split}
\hat{v}_{z}^{N+\frac{1}{4}} = &\cos(\omega_B\Delta t)\hat{v}_{z}^{N}
+ \Delta t\,q_1(\omega_B\Delta t)\left[(g/\bar{\theta})\hat{\theta}^N+\hat{\mathfrak{N}}_{z}\right] \\
&+ \tfrac{1}{2}\Delta t^2\,q_2(\omega_B\Delta t)(g/\bar{\theta}) \left[\hat{\mathfrak{M}}\right],
\end{split}\\
\begin{split}
\hat{\theta}^{N+\frac{1}{2}} = &\cos(\omega_B\Delta t)\hat{\theta}^{N}
+ \Delta t\,q_1(\omega_B\Delta t)\left[(-d\bar{\theta}/dz)\hat{v}_z^N+\hat{\mathfrak{M}}\right] \\
&+ \tfrac{1}{2}\Delta t^2\,q_2(\omega_B\Delta t)(-d\bar{\theta}/dz) \left[\hat{\mathfrak{N}}_z-\partial\hat{h}^N/\partial z\right],
\end{split}
\end{align}
\end{subequations}
where the functions $q_1$ and $q_2$ are defined:
\begin{subequations}
\begin{align}
q_1(\tau) &= \sin(\tau)/\tau,\\
q_2(\tau) &= 2[1-\cos(\tau)]/\tau^2.
\end{align}
\end{subequations}
The time differencing error for one step scales as $\Delta t^3$.  For $\omega_{C/B}\Delta t\ll 1$, $q_1$ and $q_2$ can be replaced with unity without
changing how the error scales with the size of the step.

\subsubsection{Step 1: Advection --- Explicit treatment of Coriolis and buoyancy forces with gravity-wave damping}

As an internal gravity wave propagates from a high density region into a low 
density region, its amplitude increases so as to conserve energy flux.  If
the density contrast is large, velocity and thermodynamic fluctuations can
become sufficiently large so as to invalidate the anelastic approximation
and/or violate the CFL condition.  We have developed a strategy to control
gravity waves that is akin to ``negative feedback'' in circuit theory. 
Mathematically, the origin of the buoyant oscillations is the fact that the
time derivative of the vertical velocity depends linearly on the potential
temperature, and vice versa.  Physically,  a hot parcel of fluid rises into
less dense material, expands and cools; it then sinks, compresses, and
heats-up, and the cycle continues.  To damp these oscillations, we add a term
proportional to the \textit{time derivative} of the potential temperature to
the vertical velocity equation, and vice versa.  This introduces a term that
is proportional to $-\tilde{v}_z$ into the equation for the vertical
velocity, and a term that is proportional to $-\tilde{\theta}$ into the
equation for the potential temperature.  In the following integration scheme,
we add the Coriolis and buoyancy terms to the nonlinear advection terms and
treat them all with an Adams-Bashforth scheme:
\begin{subequations}
\begin{align}
\begin{split}
\boldsymbol{\hat{\mathfrak{N}}} \equiv & \frac{3}{2}\left[\widehat{\left(\boldsymbol{\tilde{v}\times\tilde{\omega}}\right)}^{N}-2\Omega\boldsymbol{\hat{z}\times\hat{v}}^{N}+\sigma\hat{v}_y^{N}\boldsymbol{\hat{x}}+\frac{\hat{\theta}^{N}}{\bar{\theta}}g\boldsymbol{\hat{z}}\right]\\
-&\frac{1}{2}\left[\widehat{\left(\boldsymbol{\tilde{v}\times\tilde{\omega}}\right)}^{N-1}-2\Omega\boldsymbol{\hat{z}\times\hat{v}}^{N-1}+\sigma\hat{v}_y^{N-1}\boldsymbol{\hat{x}}+\frac{\hat{\theta}^{N-1}}{\bar{\theta}}g\boldsymbol{\hat{z}}\right],
\end{split}\\
\begin{split}
\hat{\mathfrak{M}} \equiv & \frac{3}{2}\left[-\widehat{\left(\boldsymbol{\tilde{v}\cdot\nabla}\tilde{\theta}\right)}^{N}-\left(\frac{d\bar{\theta}}{dz}\right)\hat{v}_z^{N}\right]\\
-&\frac{1}{2}\left[-\widehat{\left(\boldsymbol{\tilde{v}\cdot\nabla}\tilde{\theta}\right)}^{N-1}-\left(\frac{d\bar{\theta}}{dz}\right)\hat{v}_z^{N-1}\right].
\end{split}
\end{align}
\end{subequations}
The velocity and potential temperature are updated via:
\begin{subequations}
\begin{align}
\boldsymbol{\hat{v}}^{N+\frac{1}{4}} &= \boldsymbol{\hat{v}}^{N} + \Delta t\left[\boldsymbol{\mathfrak{N}}
+\boldsymbol{\hat{z}}\beta(z)\left(\frac{g}{\bar{\theta}\omega_B}\right)\mathfrak{M}\right],\\
\hat{\theta}^{N+\frac{1}{2}} &= \hat{\theta}^{N} + \Delta t\left[\mathfrak{M}-\beta(z)\left(\frac{\bar{\theta}\omega_B}{g}\right)\left(\mathfrak{N}_z-
\frac{\partial\hat{h}^{N}}{\partial z}\right)
\right].
\end{align}
\end{subequations}
The function $\beta(z)$ determines the range of $z$ at which
the damping operates; $\beta(z)=0$ for most of the domain, but
will be nonzero and small in the low density regions.

We note that one could damp buoyant oscillations just by adding a term
proportional to $-\tilde{v}_z$ (instead of $\partial\tilde{\theta}/\partial t$) to the vertical velocity equation,  and/or a term proportional to $-\tilde{\theta}$ (instead of $-\partial\tilde{v}_z/\partial t$) to the potential
temperature equation.  However, the advantage of adding time derivatives is
that as the flow approaches a steady-state, the damping naturally turns off.
We will explore the effects of this gravity-wave damping mechanism in
Section~4.3.

\subsubsection{Step 2: Hyperviscosity}

Unlike finite-difference codes which have a ``grid viscosity'' associated
with the spatial differencing, spectral codes have no inherent grid
dissipation.  Energy will cascade down (via nonlinear interactions) to the
smallest resolved scales (highest wavenumbers) where it will accumulate and
eventually cause large truncation and aliasing errors.  To control this,
artificial viscosity or some other kind of low-pass filtering is applied to
dissipate the energy at the smallest scales \citep{canuto88,boyd98}.
This method is related to ``large-eddy'' simulations in which the large
scale motions are fully resolved, whereas the small scale flow is treated
with a sub-grid scale turbulence model.

For the two horizontal Fourier directions, we apply a hyperviscosity of the
form: $\nu_{\bot}^{hyp}(-1)^{p+1}\hat{\nabla}_{\bot}^{2p}$, where
$\nu_{\bot}^{hyp}$ is a hyperviscosity coefficient, and $p$ is an integer
between 1 and 6.  This is simple to implement because the hyperviscosity
operator is an exact eigenoperator of the Fourier basis functions:
\begin{equation}
(-1)^{p+1}\hat{\nabla}_{\bot}^{2p}\exp(ik_xx+ik_yy) = -(k^2_x+k^2_y)^p\exp(ik_xx+ik_yy).
\end{equation}
The vertical direction is not resolved with a Fourier basis and is not
periodic.  Applying a real diffusion operator to the vertical direction is
problematic because the second derivative operator (or its powers) is not an
eigenoperator of the Chebyshev polynomial or rational Chebyshev function
bases.  Furthermore, a real diffusion operator would raise the order of the
differential equations and require additional unphysical vertical boundary
conditions and/or create spurious boundary layers.  \citet{boyd98} advocates 
replacing the second derivative operator with an eigenoperator of the basis
functions.  For example, for the Chebyshev polynomial basis (where we assume
$L_z=1$):
\begin{subequations}
\begin{align}
\frac{d^2}{dz^2}\longrightarrow\sqrt{1-z^2}\frac{d}{dz}\left[\sqrt{1-z^2}\frac{d}{dz}\right],\\
\sqrt{1-z^2}\frac{d}{dz}\left[\sqrt{1-z^2}\frac{d}{dz}\right]T_n(z) = -n^2T_n(z).
\end{align}
\end{subequations}
Likewise, there is a similar replacement for the eigenoperator of the
rational Chebyshev function basis.  The chief advantage is that because the
eigenoperators are singular at the boundaries, no additional unphysical
boundary conditions are needed and no spurious boundary layers are created.

In practice, we apply hyperviscosity every time step:
\begin{subequations}
\begin{align}
f_{hyp}&\equiv \exp[-\Delta t(\nu_{\bot}^{hyp}k_{\bot}^{2p}+\nu_z^{hyp}n^{2p})],\\
\boldsymbol{\hat{v}}^{N+\frac{2}{4}} &= \boldsymbol{\hat{v}}^{N+\frac{1}{4}}\cdot f_{hyp},\\
\hat{\theta}^{N+1} &= \hat{\theta}^{N+\frac{1}{2}}\cdot f_{hyp},
\end{align}
\end{subequations}
where $k_{\bot}^2\equiv k_x^{'2} + (k'_y+\sigma t k'_x)^2$ and $n$ is
the order of the Chebyshev polynomial or rational Chebyshev function.
The hyperviscosity coefficients are initially set to values such that the
highest resolved Fourier or Chebyshev number has an $e$-folding
time equal to one timestep.  They can be dynamically adjusted every few
hundred timesteps so that the energy spectrum does not curl-up at the highest
wavenumbers.

\subsubsection{Step 3: Poisson-like equation for pressure/enthalpy}

Subtraction of the enthalpy gradient is done with a semi-implicit
Crank-Nicholson (globally second-order in time) algorithm:
\begin{subequations}\label{E:crank_nicholson}
\begin{align}
\boldsymbol{\hat{v}}^{N+\frac{3}{4}} &= \boldsymbol{\hat{v}}^{N+\frac{2}{4}}-
\tfrac{1}{2}\Delta t \boldsymbol{\hat{\nabla}}^N\hat{h}^{N},\label{E:crank_nicholson1}\\
\boldsymbol{\hat{v}}^{N+1} &= \boldsymbol{\hat{v}}^{N+\frac{3}{4}}-
\tfrac{1}{2}\Delta t \boldsymbol{\hat{\nabla}}^{N+1}\hat{h}^{N+1},\label{E:crank_nicholson2}
\end{align}
\end{subequations}
where $\boldsymbol{\hat{\nabla}}^N\equiv[ik'_x,i(k'_y+\sigma t^Nk'_x),\partial/\partial z]$.  Step \eqref{E:crank_nicholson1}, the explicit step, uses the
current enthalpy, whereas step \eqref{E:crank_nicholson2}, the implicit step,
uses the future enthalpy.  We require that the final velocity satisfy the
anelastic constraint, which yields a Poisson-like equation for the final
enthalpy:
\begin{subequations}\label{E:Poisson}
\begin{align}
\left[\boldsymbol{\hat{\nabla}}^{N+1}+\left(\frac{d\ln\bar{\rho}}{dz}\right)\boldsymbol{\hat{z}}\right]\cdot\boldsymbol{\hat{v}}^{N+1}&=0,\label{E:Poisson1}\\
\left[\frac{\partial^2}{\partial z^2} + \left(\frac{d\ln\bar{\rho}}{dz}\right)\frac{\partial}{\partial z} - (k_{\bot}^{N+1})^2\right]\hat{h}^{N+1}
&= \frac{2}{\Delta t}\left[\boldsymbol{\hat{\nabla}}^{N+1}+\left(\frac{d\ln\bar{\rho}}{dz}\right)\boldsymbol{\hat{z}}\right]\cdot\boldsymbol{\hat{v}}^{N+\frac{3}{4}},
\label{E:Poisson2}
\end{align}
\end{subequations}
where $(k_{\bot}^{N+1})^2\equiv k_x^{'2} + (k'_y+\sigma t^{N+1} k'_x)^2$.

If the vertical domain is finite (\ie Chebyshev polynomial basis), then
explicit boundary conditions will be needed to solve this Poisson-like
equation.  We require that the vertical velocity vanish at the walls.
Imposing this on the $z$-component of \eqref{E:crank_nicholson2} yields: 
\begin{subequations}\label{E:boundary_conditions}
\begin{align}
\hat{v}_{z}^{N+1}\Bigl\vert_{z=\pm L_z}&=0, \label{E:boundary_condition1}\\
\frac{\partial\hat{h}}{\partial z}^{N+1}\Bigl\vert_{z=\pm L_z} &= \frac{2}{\Delta t}\hat{v}_{z}^{N+\frac{3}{4}}\Bigl\vert_{z=\pm L_z}.
\label{E:boundary_condition2}
\end{align}
\end{subequations}
If the vertical domain is infinite (\ie rational Chebyshev function basis),
then no explicit boundary conditions are needed to invert the Poisson-like
equation because the vertical basis functions individually satisfy the
boundary conditions (\ie Galerkin method with natural boundary conditions):
$dC_n(z)/dz \rightarrow 0$ and $S_n(z) \rightarrow 0$ as
$z \rightarrow \pm \infty.$

The Poisson-like equation is singular for the case of $k'_x=k'_y=0$.  The
anelastic condition for this mode implies that there can be no net vertical
momentum across any horizontal plane:
\begin{subequations}
\begin{align}
\hat{v}_{z,00}^{N+1} &= 0,\\
\frac{\partial\hat{h}^{N+1}_{00}}{\partial z} &= \frac{2}{\Delta t}\hat{v}_{z,00}^{N+\frac{3}{4}}.
\end{align}
\end{subequations}
The integration constant for $\hat{h}^{N+1}$ is determined by the condition
that the total mass in the domain is conserved.

One can write the Poisson-like equation \eqref{E:Poisson2} in matrix form:
$\boldsymbol{\mathbb{M}}\boldsymbol{\hat{h}}=\boldsymbol{\hat{g}}$,
where $\boldsymbol{\mathbb{M}}$ is a matrix containing the Laplacian-like
differential operator, $\boldsymbol{\hat{h}}$ is a column vector of vertical
spectral coefficients for the enthalpy, and $\boldsymbol{\hat{g}}$ is a
column vector of vertical spectral coefficients for the right-hand side of
\eqref{E:Poisson2}.  When needed, the last two rows of the matrix
$\boldsymbol{\mathbb{M}}$ are overwritten with the boundary condition data
\eqref{E:boundary_condition2}, at the expense that the divergence condition
will not be satisfied for the two highest vertical spectral modes.  This is
usually an exponentially small error, but if it is undesirable, one can use
the tau-method with Greens functions to correct the divergence at the two
highest spectral modes (see Appendix B; \citealp{marcus84a}).

The speed and efficiency in inverting the Poisson-like equation depends on
the functional form of the stratification term $d\ln\bar{\rho}/dz$.  If the
stratification term is particularly simple (\eg zero for no stratification,
a constant for constant stratification, or proportional to height $z$ as in
a protoplanetary disk), then one can use Chebyshev or rational Chebyshev
recursion relations; the matrices will be banded and consequently quick to
invert.  In particular, for the finite domain with the Chebyshev polynomial
basis, the matrix equation can be rewritten as:
$\boldsymbol{\mathbb{A}\hat{h}} = \boldsymbol{\mathbb{B}\hat{g}}$, where
$\boldsymbol{\mathbb{A}}$ and $\boldsymbol{\mathbb{B}}$ are pentadiagonal
matrices (see Appendix A, \citealp{gottlieborszag77,canuto88}).  If the
stratification term is an odd function of $z$, then the even and odd
Chebyshev modes decouple, reducing the pentadiagonal matrix equation to two
tridiagonal matrix equations for the even and odd Chebyshev modes.  For the
infinite domain with rational Chebyshev function basis, the matrix
$\boldsymbol{\mathbb{M}}$ is nonadiagonal (see Appendix C).  If the
stratification term is an odd function of $z$, then the even and odd rational
Chebyshev modes decouple, reducing the nonadiagonal matrix equation to two
pentadiagonal matrix equations for the even and odd rational Chebyshev modes.
On the other hand, if the stratification term is a more complicated function 
of height, then the terms multiplied by the stratification term will have to 
be computed in physical space on a grid of collocation points, and then
transformed back to function space.  The matrices in the Poisson-like
equation will consequently be full and slower to invert.

\section{Numerical Tests}

\subsection{Parallelization \& timing analysis}

\begin{figure}
\begin{center}
\mbox{\epsfig{file=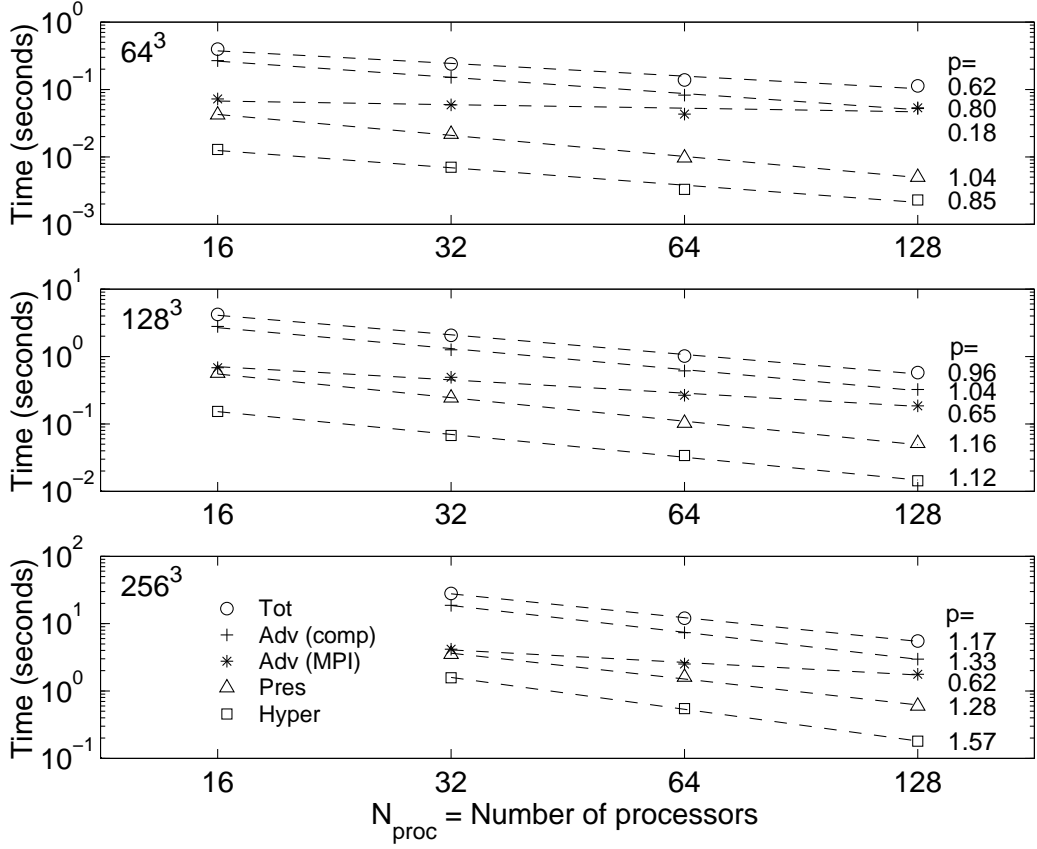,height=4.5in}}
\caption{\label{F:code_timing} Average wall-clock or real time to compute one
timestep using 16, 32, 64, and 128 processors on the IBM Blue Horizon at the
San Diego Supercomputer Center.  Parallelization is implemented with the
Message Passing Interface (MPI) protocol.  `Tot' = Total time for one
timestep, `Adv (comp)' = computation time for advection step,  `Adv (MPI)' = 
MPI communication time for advection step, `Pres' = computation time for
pressure step, `Hyper' = computation time for hyperviscosity step.  The
dashed lines are power-law fits;  the numbers to the right of the data are
the best-fit exponent: Time $\propto N_{proc}^{-p}$.  For reference, we note 
that the DataStar runs approximately two times faster than the Blue Horizon.}
\end{center}
\end{figure}

We have developed our simulation for use on parallel supercomputers such as
the Blue Horizon (decommissioned in April 2004) and DataStar supercomputers
at the San Diego Supercomputer Center.  The DataStar has 176 nodes of 8
processors (CPU speed of 1.5 GHz and peak performance of 6.0 GFlops).  Each
node shares 16GB of memory.

Different horizontal Fourier modes interact only through the nonlinear
advection terms, whereas the vertical spectral modes (Chebyshev polynomials
or rational Chebyshev functions) are fully coupled at every stage of the
computation.  This motivated us to divide the computational domain only along
the horizontal directions.  Each processor works with a small subset of
horizontal data, but all of the corresponding vertical data: in physical
space, the $j$th processor would contain data for: $x_1^j\le x<x_2^j$,
$y_1^j\le y<y_2^j$, $-L_z\le z\le+L_z$; in wavenumber space, a given
processor would contain data for: $k_{x,1}^j\le k_x<k_{x,2}^j$, $k_{y,1}^j\le k_y<k_{y,2}^j$,  $0\le n\le N_z$.  The only time different processors 
communicate with each other is when a horizontal FFT needs to be computed.
There are generally two common strategies for this: one could use a FFT that
works across processors, or one could transpose the data so that a given
direction is not split across processors.  For example, if one wanted to
compute a FFT along the $x$ direction, one would swap the data so that each
processor had a subset of the $y$ and $z$ data, but all of the corresponding
$x$ data.  We have implemented this second  strategy with an efficient
transpose algorithm developed by Alan Wray at  NASA-Ames/Stanford Center for 
Turbulence Research (Wray, personal communication).

Figure \ref{F:code_timing} shows the results of timing tests for a version of
the code that used the Chebyshev polynomial basis.  The tests were done on
the Blue Horizon and analyzed with the Vampir parallel trace tool.  For
reference, we note that the DataStar runs approximately two times faster than
the Blue Horizon. We present the average wall-clock or real time to compute
one timestep for $64^3$, $128^3$, and $256^3$ domains on 16, 32, 64, and 128 
processors.  We fit power laws ($T\propto N_{proc}^{-p}$) to the timing data 
and determined how the times for different steps in the algorithm scale with 
number of processors (ideal performance is $p\approx 1$).  As expected, the
advection step is the most time-intensive step in the algorithm 80-90\% of
the time is spent in this stage alone due to the large number of FFTs
required to compute the nonlinear terms.  The computation time (`Adv comp')
scales well with the number of processors, whereas the communication time
(`Adv MPI') has a much flatter scaling. 

\subsection{2D vortex dynamics in a shearing flow}

The simplest way to validate the shearing box algorithm is to confirm that
the code correctly simulates 2D vortex dynamics.  \citet{ms71} analytically
determined the steady state solutions for 2D elliptical vortices of uniform
vorticity $\tilde{\omega}_z$ embedded in a uniform shear of strength
$\sigma$, and showed that the aspect ratio of a vortex, $\chi\equiv\Lambda_x/\Lambda_y$, where $\Lambda_x$ and $\Lambda_y$ are the major and minor axes of
the ellipse, was determined by the ratio of the vorticity to the shear:
\begin{equation} \label{E:moore_saffman}
\frac{\tilde{\omega}_z}{\sigma} = \left(\frac{\chi+1}{\chi-1}\right)\frac{1}{\chi},
\end{equation}
which is valid for vortices that rotate in the same sense as the shear.  As
one would expect, stronger vortices are more round and compact, whereas
weaker vortices are more elongated.  \citet{kida81} showed that if such
a vortex was perturbed, its major axis would oscillate about the $x$-axis,
and its aspect ratio would oscillate about its mean value (area remaining
constant).  These motions would be damped via the occasional shedding of thin
filaments of vorticity which would eventually be dissipated by viscosity.

Figure \ref{F:2D_vortex} shows the kinetic, shear, and total energies as a
function of time for the evolution of a 2D elliptical vortex. The parameters 
for these runs were: $(L_x,L_y)=(4,2)$, $(N_x,N_y)=(256,128)$, $\Omega=1$,
$\sigma=-1.5$, $t_{rm} = 4/3$,  $\Delta t = t_{rm}/100$,
$(\Lambda_x,\Lambda_y)=(1,0.25)$, $\tilde{\omega}_z(0) = -0.625$.
The initial vortex was surrounded with a weak halo of oppositely-oriented
vorticity so that there was no net circulation around the vortex
($\Gamma\equiv\int_A\tilde{\omega}_z\,dxdy=0$).  The Kida oscillations are
readily apparent in the energy plots as the larger amplitude, longer period
oscillations ($\approx 8$ re-map periods).  As expected, increasing the
hyperviscosity resulted in the damping of these oscillations; the final state
is a steady-state vortex whose major axis is aligned with the $x$-axis and
whose aspect ratio and area are constant.

Superposed on the Kida oscillations is another low amplitude, short period
oscillation, which is most clearly seen in the plots of total energy.  This
period is exactly the re-map period and is due to interactions of the vortex
with its periodic images, which are realigned with the primary vortex every
re-map period (see Fig. \ref{F:shearing_coords}).  The amplitude of these
oscillations decreases with domain size.

\begin{figure}
\begin{center}
\mbox{\epsfig{file=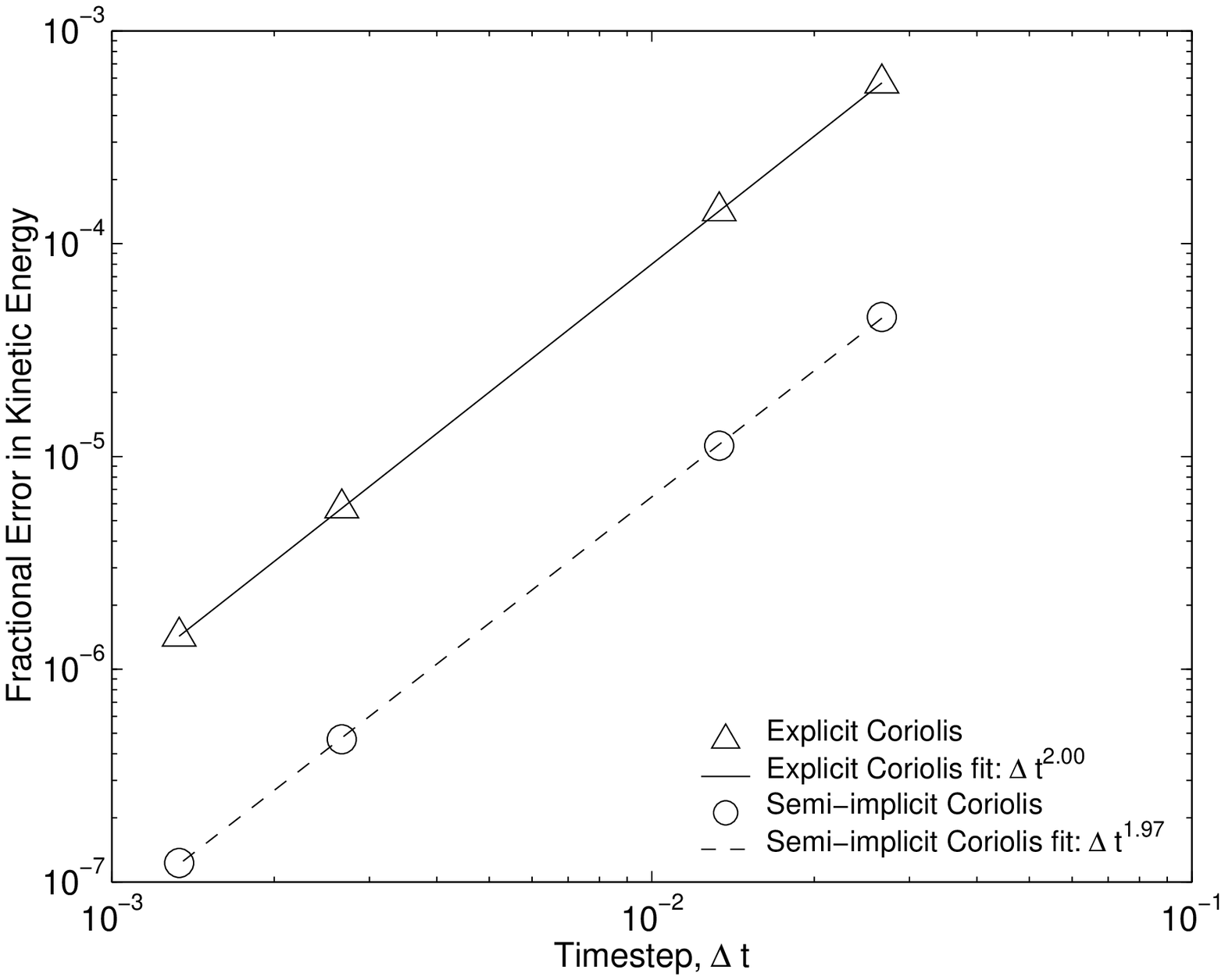,height=4in}}
\caption{\label{F:time_int_error_2D} Fractional error in final kinetic energy
after one re-map period versus timestep size for explicit and semi-implicit
treatment of the Coriolis force.  Both methods are second-order accurate, but
the error is an order of magnitude smaller for the semi-implicit method.}
\end{center}
\end{figure}

In Figure \ref{F:time_int_error_2D}, we plot the percent error in kinetic
energy after one re-map period as a function of timestep size for the
explicit and semi-implicit treatments of the Coriolis (modified by shear)
terms.  We fit power-laws and confirmed that both schemes are second-order
accurate in time.  The semi-implicit treatment, however, has an error that is
a order of magnitude smaller than the explicit scheme.

\setlength{\unitlength}{1 in}
\begin{figure}
\begin{center}
\begin{picture}(6,3)
\put(0,0){\epsfig{file=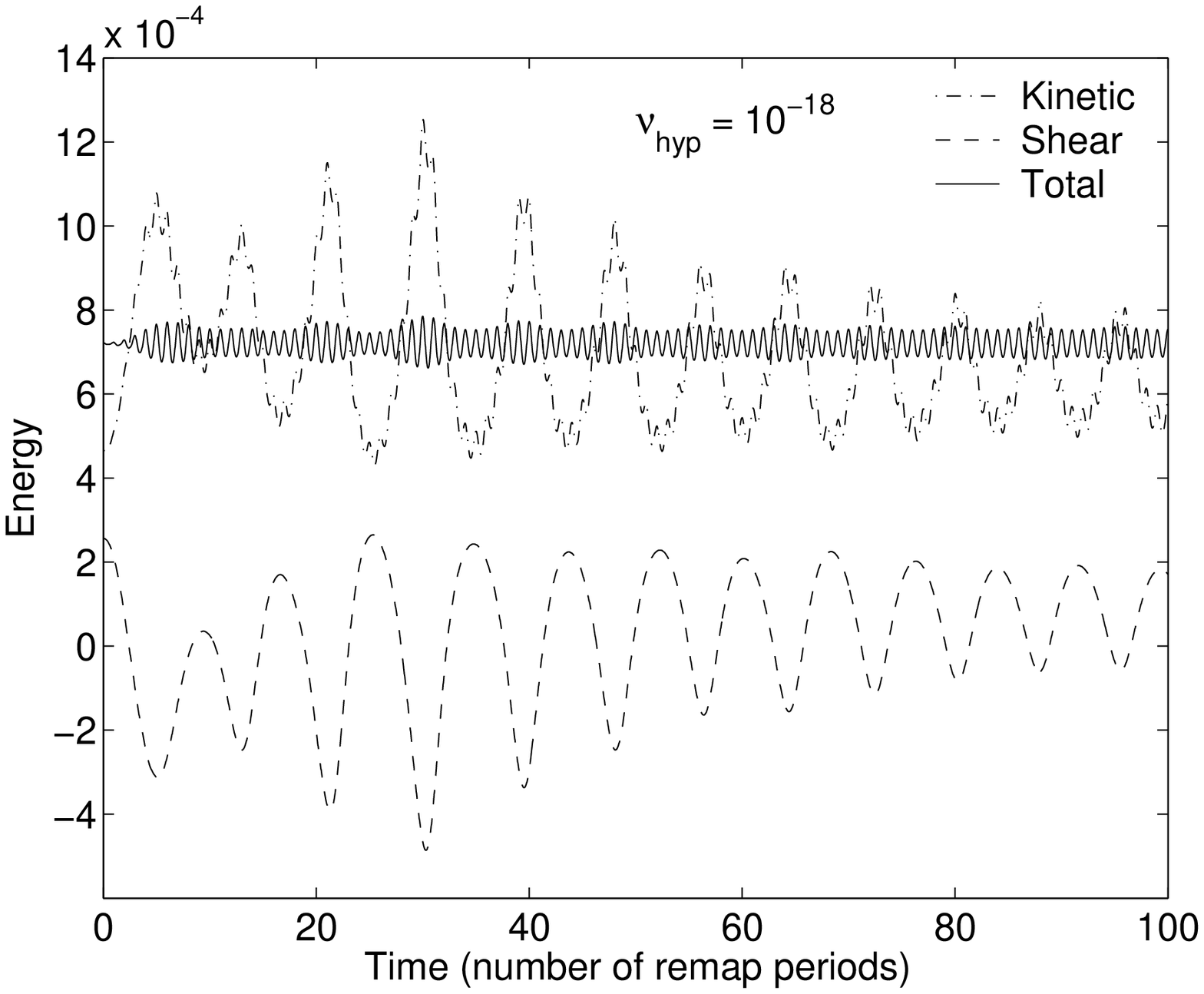,width=3 in}}
\put(3,0){\epsfig{file=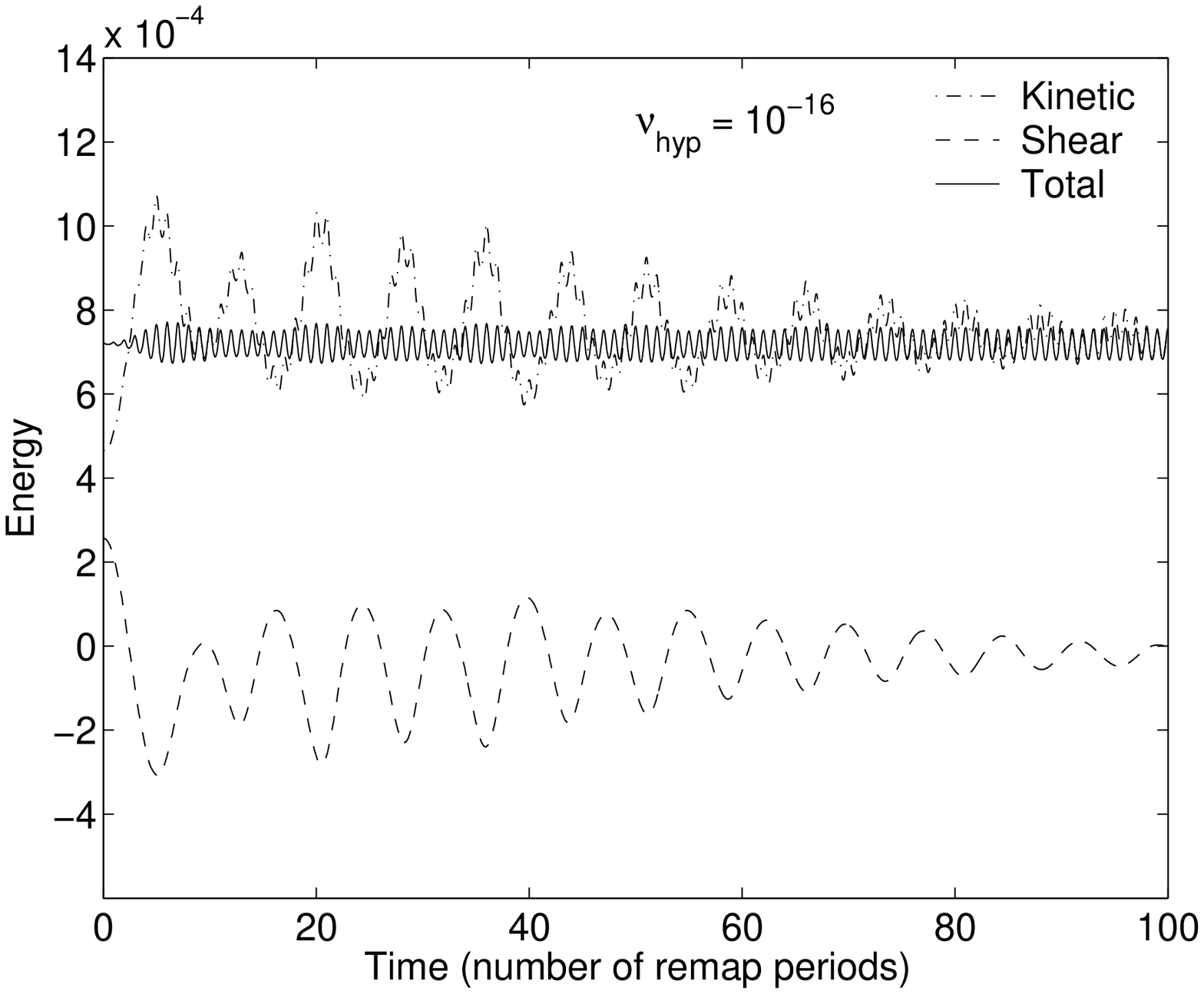,width=3 in}}
\end{picture}
\caption{\label{F:2D_vortex} Kinetic, shear, and total energy for 2D
elliptical vortex embedded in a linear shear.  The larger amplitude, longer
period ($\approx 8$ re-map periods) oscillations are associated with the
major axis of the vortex oscillating about the $x$-axis and with the aspect
ratio oscillating around its mean value, as described by \citet{kida81}.  The
lower-amplitude, shorter period oscillations (exactly equal to the re-map
period) are due to interactions with the periodic image vortices, which are
realigned with the primary vortex every re-map period.  The Kida oscillations
are damped via the shedding of thin filaments which are dissipated by
hyperviscosity.}
\end{center}
\end{figure}

\subsection{Linear eigenmodes in a stratified protoplanetary disk}

To test the vertical dimension of the code, we investigate the behavior of
linear eigenmodes (\ie internal gravity waves) in a stratified background.
The background shear makes it difficult to compute eigenmodes because
the linearized equations depend explicitly on $\sigma y\partial/\partial x$
in the original coordinates, or explicitly on $\sigma tik_x$ in the shearing
coordinates.  If we restrict  the following analysis to the case of
$\partial/\partial x = ik_x = 0$, then the process of finding eigenmodes is
straight-forward.  We look for eigenmodes of the form
$\hat{q}(z)\exp(ik_yy-i\omega t)$, where $q$ represents any variable.  The
real part of the eigenvalue is related to the phase speed of the wave,
$c=\mathcal{RE}(\omega)/k_y$, whereas the imaginary part
$\gamma=\mathcal{IM}(\omega)$ corresponds to wave growth (if positive) or
damping (if negative).  The linearized equations are:
\begin{subequations}
\begin{align}
-i\omega\hat{v}_x &= (2\Omega+\sigma)\hat{v}_y,\\
-i\omega\hat{v}_y &= -2\Omega\hat{v}_x -ik_y\hat{h},\\
-i\omega\hat{v}_z &= -d\hat{h}/dz + (\hat{\theta}/\bar{\theta})g
-\beta\omega_B\hat{v}_z,\\
-i\omega\hat{\theta} &= -(d\bar{\theta}/dz)\hat{v}_z
-\beta\omega_B\left[\hat{\theta} - (\bar{\theta}/g) d\hat{h}/dz\right],\\
0 &= ik_y\hat{v}_y + d\hat{v}_z/dz + (d\ln\bar{\rho}/dz)\hat{v}_z,
\end{align}
\end{subequations}
where we have included the damping terms described in section 3.3.2.
If the damping terms are neglected, then this system of equations can be
simplified to the second-order eigenvalue problem for the vertical
velocity fluctuation:
\begin{equation}
\left(\frac{\omega}{\omega_C}\right)^2\left[\mathcal{L}-k_y^2\right]\hat{v}_z =
\left[\mathcal{L}-k_y^2\left(\frac{\omega_B}{\omega_C}\right)^2\right]\hat{v}_z,
\end{equation}
where we have defined the second-order linear differential operator:
\begin{equation}
\mathcal{L}f \equiv \frac{d}{dz}\left[\frac{df}{dz} + \left(\frac{d\ln\bar{\rho}}{dz}\right)f\right].
\end{equation}
If the eigenvalue problem is solved on the finite domain with the Chebyshev
polynomial basis, then boundary conditions must be imposed:
$\hat{v}_z(z\!=\!\pm L_z)=0$.  On the infinite domain with the rational
Chebyshev function basis, boundary conditions are naturally satisfied by the 
basis functions (\ie Galerkin method).

In order to proceed, we need to specify the background state.  Here, we
choose conditions relevant to a protoplanetary disk of gas and dust in orbit 
around a newly-formed protostar \citep{barranco00a, barranco05a}.  The
velocity profile of the gas is very nearly Keplerian:
$V_K(r)\equiv r\Omega_K(r)\equiv\sqrt{GM_{\star}/r}$, where $G$ is the
gravitational constant, $M_{\star}$ is the mass of the protostar, and $r$ is 
the cylindrical radius to the protostar.  The horizontal shear rate of this
base flow is: $\sigma_K\equiv r(\partial\Omega_K/\partial r)=-(3/2)\Omega_K.$
Keplerian shear is anticyclonic (as indicated by the negative sign), and is
comparable in magnitude to the rotation rate itself.  We do not simulate the 
entire disk; rather, we we simulate the hydrodynamics only within a small
patch $(\Delta r\ll r_0,\Delta\phi\ll 2\pi)$ that co-rotates with the gas at 
some fiducial radius $r_0$ with angular velocity $\Omega_0\equiv\Omega(r_0)$.
We map this patch of the disk onto a Cartesian grid:  $r-r_0 \rightarrow y$, 
$r_0(\phi - \phi_0) \rightarrow -x$, $z\rightarrow z$, $v_r \rightarrow v_y$,
$v_{\phi} \rightarrow -v_x$, and $v_z\rightarrow v_z$.  In this rotating
patch, the velocity profile of the gas is very nearly a linear shear flow
with $\sigma_0\equiv\sigma_K(r_0)=-(3/2)\Omega_0$.   The Coriolis frequency
(modified by the shear) is $\omega_C=\Omega_0$.

The vertical component of the protostellar gravity in this patch of the disk 
is very nearly $g_z(z)\approx\Omega_0^2z$.  We assume that the background is
constant temperature $\bar{T}=T_0$, which yields a hydrostatic density
profile that is Gaussian:
\begin{subequations}
\begin{align}
\bar{\rho}(z) &= \rho_0\exp(-z^2/2H_0^2), \quad H_0^2\equiv\mathcal{R}T_0/\Omega_0^2,\\
H_{\rho}(z)&\equiv \vert\left(d\ln\bar{\rho}/dz\right)^{-1}\vert = H_0^2/|z|.
\end{align}
\end{subequations}
The Brunt-V\"{a}is\"{a}l\"{a} frequency is:
\begin{equation}
\omega_B(z) = \sqrt{\mathcal{R}/C_P}\,\Omega_0|z|/H_0.
\end{equation}
The disk can be divided into two regimes: a nearly unstratified midplane
($\omega_B<\Omega_0$, $|z|/H_0\lesssim 1.6$) and a strongly stratified
atmosphere above and below the midplane
($\omega_B>\Omega_0$, $|z|/H_0\gtrsim 1.6$).

\begin{figure}
\begin{center}
\mbox{\epsfig{file=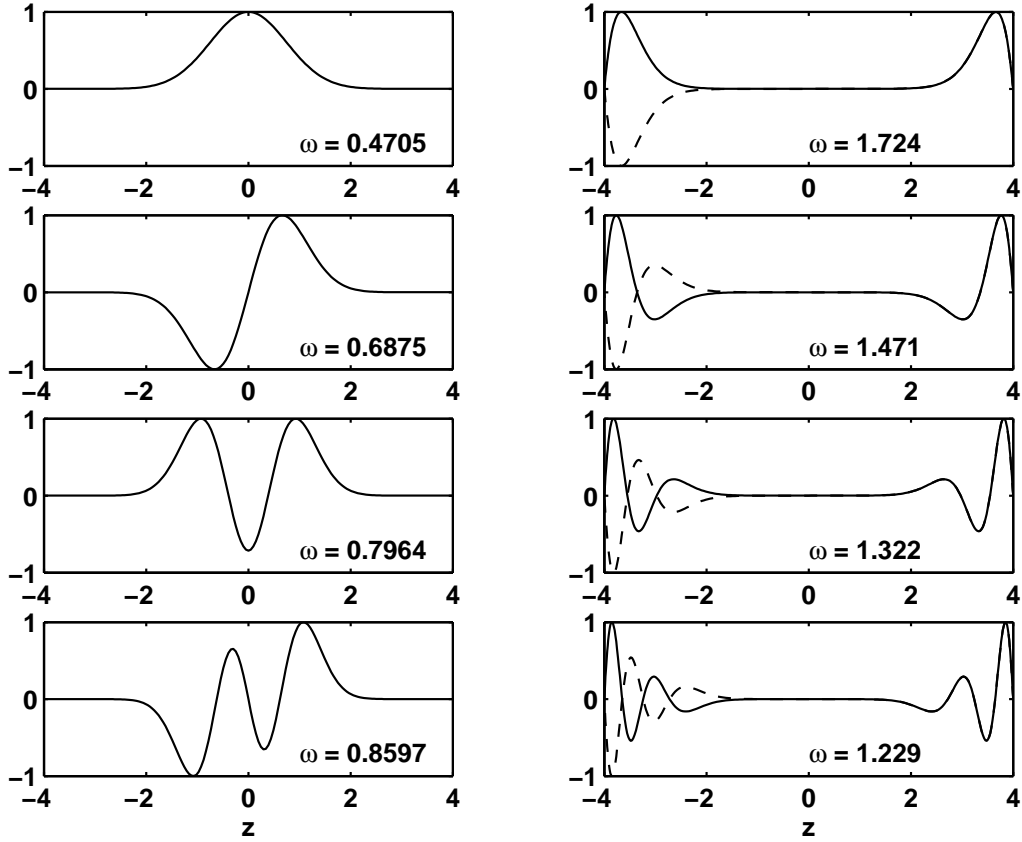,height=4.5in}}
\caption{\label{F:gravity_wave_plots}
Vertical velocity component of the undamped ($\beta=0$) eigenmodes in a
stratified protoplanetary disk for $(k_x,k_y) =(0,\pi/H_0)$.  There are two
kinds of eigenmodes: those localized around the weakly stratified midplane
with frequencies $\omega<\Omega_0$, and those localized 
in the strongly stratified regions near the boundaries with frequencies
$\omega>\Omega_0$. }
\end{center}
\end{figure}

Figure \ref{F:gravity_wave_plots} shows the vertical velocity component of
the undamped ($\beta=0$) eigenmodes for $k_yH_0=\pi$.  The eigenvalues are
purely real.  There are two kinds of eigenmodes: those confined to the weakly
stratified midplane of the disk with frequencies $\omega<\Omega_0$, and those
confined to the strongly stratified regions near the boundaries with
frequencies $\omega>\Omega_0$.  The latter class of eigenmodes do not exist
when the domain is infinite (rather, there is a continuous spectrum of waves
whose velocity amplitudes are formally unbounded as $|z|\rightarrow\infty$). 

\begin{figure}
\begin{center}
\mbox{\epsfig{file=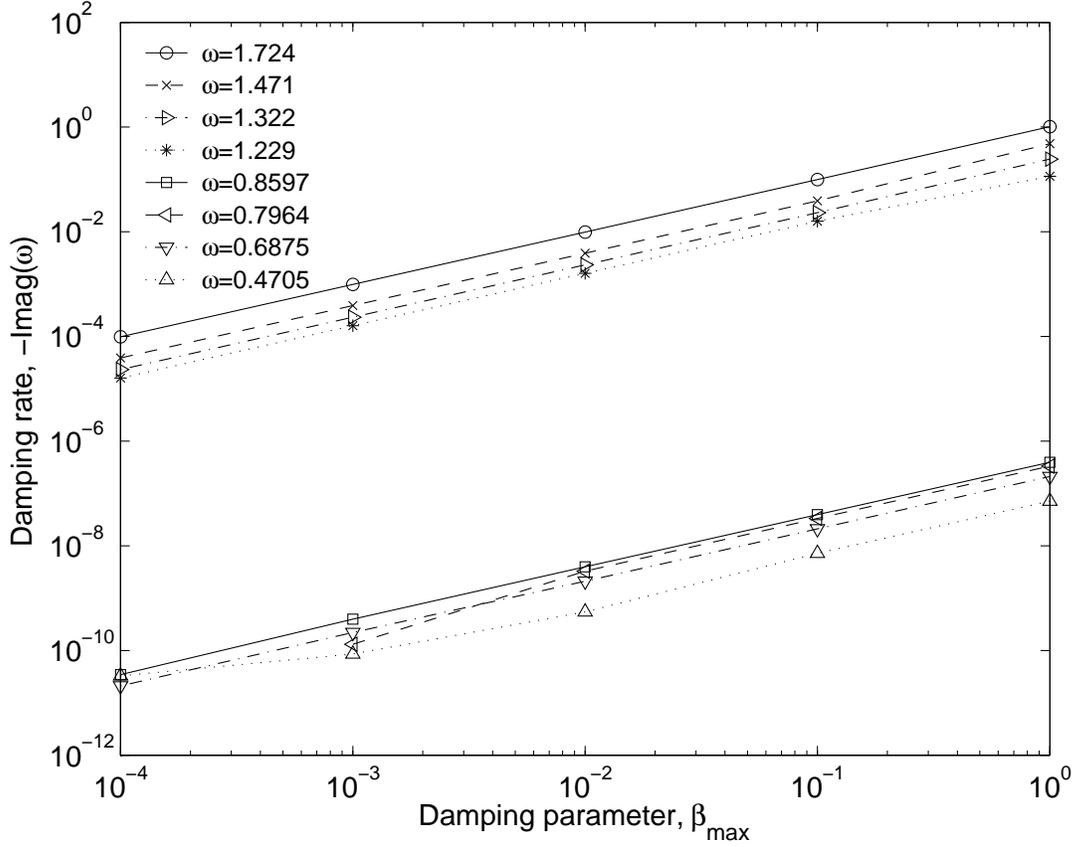,height=4.5in}}
\caption{\label{F:gravity_wave_damping}
Effect of damping on linear eigenmodes.  Eigenmodes that are localized near
the midplane (lower group) are weakly damped, whereas eigenmodes that are
localized far from the midplane (upper group) are strongly damped.  Damping
rate is linearly proportional to $\beta_{max}$.}
\end{center}
\end{figure}

To investigate the effect of damping on these eigenmodes, we need to specify
a profile for $\beta(z)$, for example:
\begin{equation}
\beta(z) = \beta_{max}\{1 + \tanh[a(z-z_c)/H_0] + \tanh[a(z_c-z)/H_0]\},
\end{equation}
which is nearly zero for most of the domain, but rises sharply at
$|z|\approx z_c$ to its maximum value $\beta_{max}$; the parameter $a$ sets
how steep the rise is.  In the calculations presented here, we take $a=4$ and
$z_c=3H_0$.  Table \ref{T:eigenvalues} and Fig.\ref{F:gravity_wave_damping}
show the effect of varying $\beta_{max}$ on the two kinds of eigenmodes.
Those eigenmodes that are localized around the midplane are only weakly
damped, whereas the eigenmodes that are localized in the strongly stratified
regions near the boundary are strongly damped. 

In fully nonlinear 3D calculations, this damping is necessary for numerical
stability.  As internal gravity waves propagate from a high density regions
into low density regions, their amplitudes increase so as to conserve energy
flux, causing velocity and thermodynamic fluctuations that can become
sufficiently large so as to invalidate the anelastic approximation and/or
violate the CFL condition.

\begin{table}
\begin{tabular}{lll} \hline
$\beta_{max}$                           & $\omega=0.8597$             & $\omega=1.229$ \\
\hline
$0$            & $0.8597$                                            & $1.229$\\
$10^{-4}$ & $0.8597 - i\,3.446\times 10^{-11}$ & $1.229 - i\,1.601\times 10^{-5}$\\
$10^{-3}$ & $0.8597 - i\,3.956\times 10^{-10}$ & $1.229 - i\,1.601\times 10^{-4} $\\
$10^{-2}$ & $0.8597 - i\,3.975\times 10^{-9  }$ & $1.229 - i\,1.601\times 10^{-3} $\\
$10^{-1}$ & $0.8597 - i\,3.946\times 10^{-8  }$ & $1.231 - i\,1.600\times 10^{-2}$\\
$1           $ & $0.8597 - i\,3.945\times 10^{-7  }$ & $1.302 - i\,1.148\times 10^{-1}$\\
\hline
\end{tabular}
\caption{Effect of damping on linear eigenmodes.  An eigenmode that is
localized near the midplane (\eg $\omega=0.8597$) is weakly damped, whereas
an eigenmode that is localized far from the midplane (\eg $\omega=1.229$) is 
strongly damped.  Damping rate is linearly proportional to $\beta_{max}$.}
\label{T:eigenvalues}
\end{table}

\subsection{Merger of two 3D vortices in a protoplanetary disk}

As a preview of potential applications, we present the simulation of a merger
of two 3D vortices in the midplane of a stratified protoplanetary disk.  A
more complete discussion of simulations of vortices in protoplanetary disks
can be found in \citet{barranco00a,barranco05a}.  We use the 2D vortex
solution of \citet{ms71} and \citet{kida81} to initialize an elliptical
vortex in the midplane at $z=0$.  The vorticity is extended off the midplane 
according to a Gaussian profile:
\begin{equation}\label{E:Gaussian_vortex}
\tilde{\omega}_z(z) = \tilde{\omega}_z(0)e^{-z^2/2H_0^2}.
\end{equation}
We initialize the temperature so that the buoyancy force exactly balances
the vertical pressure force:
\begin{equation}
\tilde{T} = \frac{T_0}{\Omega_0^2z}\frac{\partial(\tilde{p}/\bar{\rho})}{\partial z}.
\end{equation}
Of course, only under very special circumstances would this procedure just so
happen to also exactly balance the temperature equation.  In general,
$\partial\tilde{T}/\partial t\ne 0$ initially, leading to an immediate
evolution of the temperature.  Vertical motions will then be generated by the
temperature changes through the buoyancy force,  and these vertical
velocities will then couple the horizontal motions at different heights.  We 
allow such a 3D vortex to evolve and relax to a quasi-equilibrium. We then
use that vortex as a template to construct an initial condition with two
vortices destined to merge.

\begin{figure}
\begin{center}
\mbox{\epsfig{file=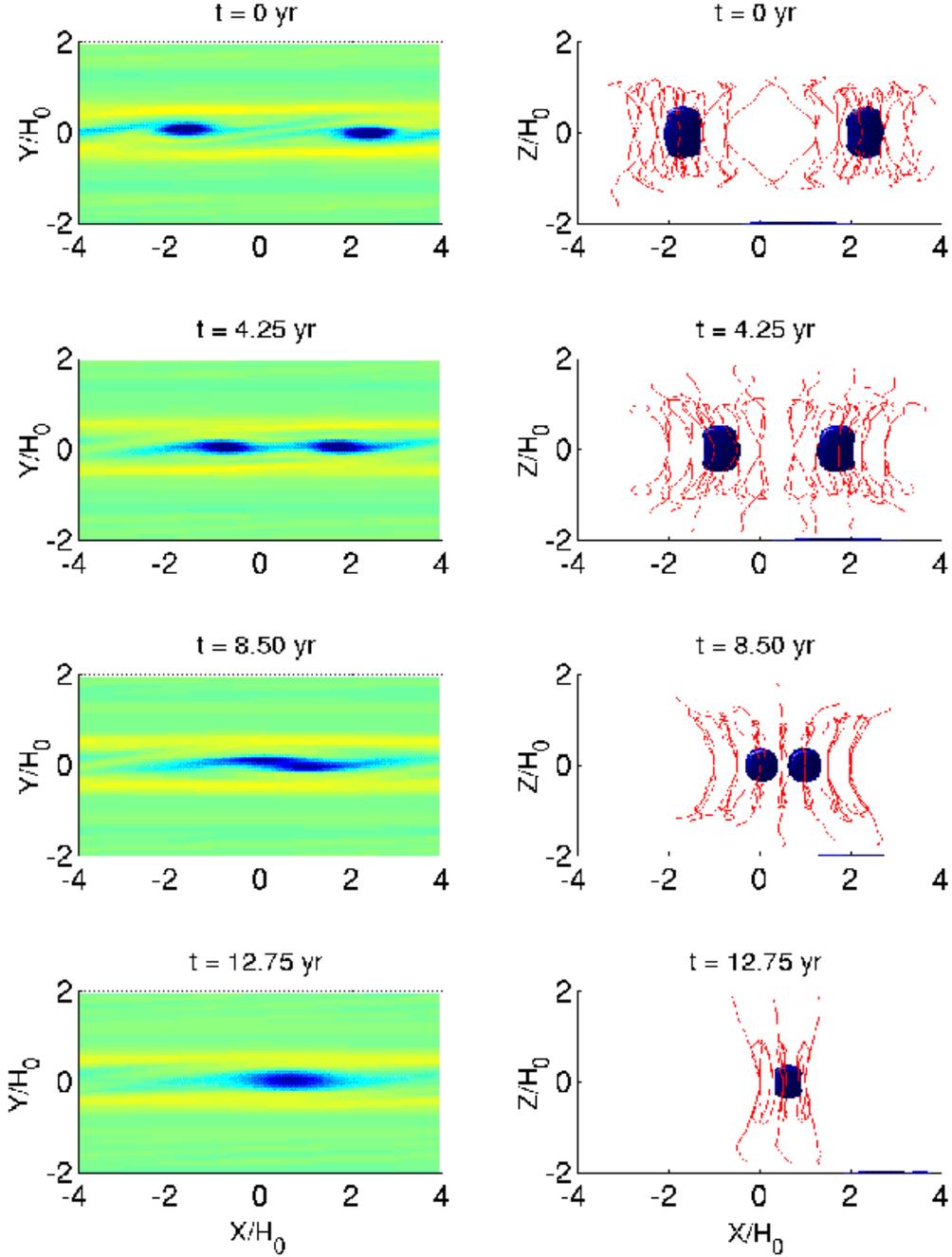,height=7in}}
\caption{\label{F:merger} The merger of two 3D vortices in the midplane of a
protoplanetary disk.  In the first column, we show $x\!-\!y$ slices in the
midplane $z\!=\!0$ of the $z$-component of vorticity.  Blue corresponds to
anticyclonic vorticity, red corresponds to cyclonic vorticity.  In the second
column, we present isovorticity surfaces for the $z$-component of vorticity
in blue, and vortex lines (lines that are everywhere tangent to the vorticity
vector) in red.  Units of time: $\tau_{orb}\equiv 2\pi/\Omega_0= 1$~yr.}
\end{center}
\end{figure}

Fig. \ref{F:merger} shows an example of a merger.  The first column shows
$x\!-\!y$ slices in the midplane $z\!=\!0$ of the $z$-component of vorticity.
Blue corresponds to anticyclonic vorticity, red corresponds to cyclonic
vorticity.  The second column shows isovorticity surfaces for the $z$
component of vorticity, in blue, and vortex lines (lines that are everywhere
tangent to the vorticity vector) in red.  The vorticity field is
divergence-free (since it is the curl of another vector field).  Thus, like
magnetic field lines, vortex lines cannot begin or end in free space, they
must either end on a boundary, extend to infinity, form closed loops, or
form unending open curves.  



\section{Conclusion}

We have developed a three-dimensional, spectral, hydrodynamic code to study
vortex dynamics in rapidly rotating, intensely sheared, and strongly
stratified systems such as giant planet atmospheres and protoplanetary disks.
The code integrates a number of specially tailored algorithms to handle the
numerical challenges associated with shear and stratification.  We used the
anelastic approximation to filter sound waves and shocks.  The horizontal
coordinates are transformed to a set of shearing coordinates that advect with
the background shear flow, and a Fourier-Fourier basis in these shearing
coordinates is used for spectral expansions in the two horizontal directions.
For the vertical direction, two different sets of basis functions have been
implemented:  (1) Chebyshev polynomials on a truncated, finite domain, and
(2) rational Chebyshev functions on an infinite domain.  Use of this latter
set is equivalent to transforming the infinite domain to a finite one with a 
cotangent mapping, and using cosine and sine expansions on the mapped
coordinate.
The nonlinear advection terms are time integrated explicitly, whereas the
enthalpy gradient and the terms responsible for internal gravity waves are
integrated semi-implicitly.  We show that gravity waves can also be damped
by adding new terms to the Euler equations.  The code is parallelized with
the Message Passing Interface (MPI), and we get excellent parallel
performance on the IBM Blue Horizon and Datastar supercomputers at the
San Diego Supercomputer Center. 

The Chebyshev polynomial code has the disadvantage that it clustered the
collocation points near unphysical walls.  This motivated us to implement the
the rational Chebyshev functions for the infinite domain.  This removed
the walls from the problem, but it did not significantly improve
the spacing of the collocation points.  Although the collocation points
were clustered more towards the center, only half of all the collocation
points were in the ``active region'' $|z|<L_z$, whereas the other half were
wasted on resolving the regions toward infinity $|z|>L_z$, where our
equations are no longer valid.  Also, the rational Chebyshev function
expansions converged poorly if any fluid dynamics occurred near the mapping
parameter $L_z$, forcing us to choose a larger value of $L_z$.  We could
improve the performance of the rational Chebyshev code if we used momentum 
ariables instead of velocity variables, since the multiplication by the mean
density made the momentum decay faster with height.


J.A.B. thanks the National Science Foundation for support via a Graduate
Student Fellowship while at Berkeley, and now with an Astronomy \& 
Astrophysics Postdoctoral Fellowship (NSF Grant AST0302146).
He also thanks the support of the Kavli Institute for Theoretical Physics
through NSF Grant PHY9907949.  P.S.M. thanks the support of NASA
Grant NAG510664 and NSF Grant AST0098465.  Computations were carried out
at the San Diego Supercomputer Center using an NPACI award.  The authors
would also like to thank Andrew Szeri, Xylar Asay-Davis, and Sushil Shetty
useful comments on the early manuscript.

\appendix
\section{Solving the Poisson-like equation for the enthalpy with Chebyshev
polynomial basis (``tau method'')}

For the case where the stratification is a linear function of height $z$, the
Poisson-like equation for the enthalpy \eqref{E:Poisson} can be written in
the form:
\begin{subequations}
\begin{align}
\mathcal{M}f(z) \equiv \left[\frac{d^2}{dz^2} - \gamma z\frac{d}{dz} - \lambda\right]f(z)&=g(z),\\
\frac{df}{dz}\Big|_{z=\pm 1}&=\chi_{\pm},
\end{align}
\end{subequations}
where we have chosen dimensions so that the boundaries are located at
$z=\pm 1$.  There will be one such equation for each pair of horizontal
wavenumbers $\{k'_x,k'_y\}$.  We expand $f(z)$ and $g(z)$ in Chebyshev
polynomial series:
\begin{equation}
f(z) = \sum_{n=0}^{N}\hat{f}_n T_n(z) \quad\text{and}\quad g(z) = \sum_{n=0}^{N}\hat{g}_n T_n(z).
\end{equation}
One can then use Chebyshev recursion relations to write the above
Poisson-like equation in matrix form:
\begin{subequations}
\begin{align}
\boldsymbol{\mathbb{M}\hat{f}} &= \boldsymbol{\hat{g}},\\
\sum_{n=1}^{N}(\pm 1)^{n}n^2\hat{f}_n &= \chi_{\pm},
\end{align}
\end{subequations}
where $\boldsymbol{\mathbb{M}}$ is a $(N\!+\!1)\times(N\!+\!1)$ matrix for
the second-order linear differential operator $\mathcal{M}$, and
$\boldsymbol{\hat{f}}$ and $\boldsymbol{\hat{g}}$ are column vectors of
$N\!+\!1$ Chebyshev spectral coefficients.  There are two problems with this
formulation: the matrix $\boldsymbol{\mathbb{M}}$ is singular and the system
is overspecified: we have $N\!+\!1$ undetermined spectral coefficients, but
$N\!+\!3$ constraints ($N\!+\!1$ equations +2 boundary conditions).  The tau
method fixes both problems simultaneously by discarding the equations for the
two highest Chebyshev modes and replacing them with the boundary conditions, 
resulting in the differential equation not being satisfied for those same two
highest modes.  The matrix $\boldsymbol{\mathbb{M}}$ is upper triangular
(except for the last two rows which now contain boundary conditions) and fast
to invert, but unfortunately is not diagonally dominant.  The largest
elements occur down the last column, and inverting it in its current form
will have large round-off errors.  \citet{gottlieborszag77} and
\citet{canuto88} show how to rewrite the Poisson-like matrix equation in the 
form:
\begin{equation}
\boldsymbol{\mathbb{A}\hat{f}} = \boldsymbol{\mathbb{B}\hat{g}},
\end{equation}
where $\boldsymbol{\mathbb{A}}$ and $\boldsymbol{\mathbb{B}}$ are diagonally
dominant, pentadiagonal matrices. \citet{gottlieborszag77} and
\citet{canuto88} give the recursion relations for the case $\gamma=0$.  Here,
we present the recursion relation for the more general case:
\begin{align}
c_{n-2}\hat{f}_{n-2}+c_{n}\hat{f}_{n}+c_{n+2}\hat{f}_{n+2} = d_{n-2}\hat{g}_{n-2}+d_{n}\hat{g}_{n}+d_{n+2}\hat{g}_{n+2},\\
\quad\text{for}\quad 2\le n\le N,\nonumber
\end{align}
where we have defined:
\begin{align}
c_{n-2} &\equiv -\frac{\alpha_{n-2}\lambda+(n-2)\gamma}{4n(n-1)}, & d_{n-2} &\equiv  \frac{\alpha_{n-2}}{4n(n-1)},\nonumber \\
c_n &\equiv 1+\frac{\beta_{n+2}(\lambda-\gamma)-(1-\beta_{n+2})\gamma(n+1)}{2(n^2-1)}, & d_n &\equiv -\frac{\beta_{n+2}}{2(n^2-1)},\nonumber \\
c_{n+2} &\equiv -\frac{\beta_{n+4}\lambda-\beta_{n+2}(2-\beta_{n+4})\gamma(n+2)}{4n(n+1)}, & d_{n+2} &\equiv \frac{\beta_{n+4}}{4n(n+1)},
\end{align}
and
\begin{equation}
\alpha_n\equiv
\begin{cases}
2 & \text{for}\quad n=0,\\
1 & \text{for}\quad n\ne 0,
\end{cases}
\quad\text{and}\quad
\beta_n\equiv
\begin{cases}
1 & \text{for}\quad n\le N,\\
0 & \text{for}\quad n >  N.
\end{cases}
\end{equation}
The boundary conditions can be rewritten:
\begin{equation}
\sum_{\substack{n=2\\\text{$n$ even}}}^{N} n^2\hat{f}_n = \frac{\chi_{+} + \chi_{-}}{2} \quad\text{and}\quad \sum_{\substack{n=1\\\text{$n$ odd}}}^{N} n^2\hat{f}_n = \frac{\chi_{+} - \chi_{-}}{2}.
\end{equation}
Note that the even and odd Chebyshev polynomials are decoupled in the matrix
equation and in the boundary conditions, so one could in fact write the
pentadiagonal system as two sets of tridiagonal systems for the even and odd 
modes. 

\section{Greens functions method to correct divergence condition for two
highest Chebyshev modes}

In Appendix A, we presented a method to solve the Poisson-like equation for
the enthalpy in a Chebyshev polynomial basis, subject to the constraint that 
the vertical velocity vanish at the two boundaries at $z=\pm 1$.  The last
two rows of the Poisson-like matrix equation were overwritten with boundary
condition data, resulting in the anelastic divergence condition not being
satisfied for the two highest Chebyshev modes.  Here, we present a method to
correct the divergence condition at the two highest modes using Greens
functions \citep{marcus84a}, at the expense that equation for the evolution
of the vertical momentum will not be satisfied at those same two highest
modes.

To make the presentation more clear, we will first define a number of
operators:
\begin{subequations}
\begin{align}
D  &\equiv (d/dz),\\
D_A&\equiv (d/dz)+(d\ln\bar{\rho}/dz),\\
\boldsymbol{\hat{\nabla}}^{N+1}&\equiv ik'_x\boldsymbol{\hat{x}} + i(k'_y+\sigma t^{N+1}k'_x)\boldsymbol{\hat{y}} + D\boldsymbol{\hat{z}},\\
\boldsymbol{\hat{\nabla}_A}^{N+1}&\equiv ik'_x\boldsymbol{\hat{x}} + i(k'_y+\sigma t^{N+1}k'_x)\boldsymbol{\hat{y}} + D_A\boldsymbol{\hat{z}},\\
\triangle_A^{N+1}&\equiv\boldsymbol{\hat{\nabla}_A}^{N+1}\cdot\boldsymbol{\hat{\nabla}}^{N+1}.
\end{align}
\end{subequations}
In the fourth fractional step \eqref{E:crank_nicholson2}, we subtract the
implicit half of the enthalpy gradient as before, but we also add two
additional functions, each consisting of exactly one Chebyshev mode:
\begin{equation}
\boldsymbol{\hat{v}}^{N+1} = \boldsymbol{\hat{v}}^{N+\frac{3}{4}} - \boldsymbol{\hat{\nabla}}^{N+1}\hat{\Pi}^{N+1} + \boldsymbol{\hat{z}}(\tau_1^{N+1} T_{M-1} + \tau_2^{N+1} T_M),
\end{equation}
where we have defined $\hat{\Pi}\equiv(\Delta t/2)\hat{h}$, and where
$\tau_1$ and $\tau_2$ are scalars (as usual, we suppress the subscripts for
wavenumber dependence).  To find the enthalpy at the $N\!+\!1$ step, we
impose the anelastic constraint
$\boldsymbol{\hat{\nabla}_A}\cdot\boldsymbol{\hat{v}}^{N+1} = 0$,
which yields:
\begin{subequations}
\begin{align}
\triangle_A^{N+1}\hat{\Pi}^{N+1} &= \boldsymbol{\hat{\nabla}_A}^{N+1}\cdot\boldsymbol{\hat{v}}^{N+\frac{3}{4}} + \tau_1^{N+1}D_AT_{M-1} + \tau_2^{N+1}D_AT_M,\\
D\hat{\Pi}^{N+1}(z\!=\!\pm 1) &= \hat{v}_z^{N+\frac{3}{4}}(z\!=\!\pm 1) + \tau_1^{N+1} T_{M-1}(\pm 1) + \tau_2^{N+1} T_M(\pm 1).
\end{align}
\end{subequations}
We can break this up into three separate Poisson-like equations:
\begin{equation}
\hat{\Pi}^{N+1} = \hat{\Pi}_0^{N+1} + \tau_1^{N+1}\Gamma_1^{N+1} + \tau_2^{N+1}\Gamma_2^{N+1}, 
\end{equation}
where $\Gamma_1$ and $\Gamma_2$ are Greens functions, and where:
\begin{subequations}
\begin{align}
\triangle_A^{N+1}\hat{\Pi}_0^{N+1} &= \boldsymbol{\hat{\nabla}_A}^{N+1}\cdot\boldsymbol{\hat{v}}^{N+\frac{3}{4}},\\
D\hat{\Pi}_0^{N+1}(z\!=\!\pm 1) &= \hat{v}_z^{N+\frac{3}{4}}(z\!=\!\pm 1),\\
\triangle_A^{N+1}\Gamma_1^{N+1} &= D_AT_{M-1},\\
D\Gamma_1^{N+1}(z\!=\!\pm 1)&=T_{M-1}(\pm 1),\\
\triangle_A^{N+1}\Gamma_2^{N+1} &= D_AT_{M},\\
D\Gamma_2^{N+1}(z\!=\!\pm 1)&=T_{M}(\pm 1).
\end{align}
\end{subequations}
Each of these 3 Poisson-like equations can be solved via the tau method as
in Appendix A.  By construction, $\boldsymbol{\hat{\nabla}_A}^{N+1}\cdot\boldsymbol{\hat{v}}^{N+1}$ equals zero for all Chebyshev modes except the two
highest; for those modes, the values of the anelastic divergence are
functions of $\tau_1^{N+1}$ and $\tau_2^{N+1}$.
We use these two degrees of freedom to impose that the anelastic divergence
equal zero at the two highest modes as well, which yields the
$2\times 2$ matrix equation:
\begin{equation}
\begin{pmatrix}\alpha_{11} & \alpha_{12}\\ \alpha_{21} & \alpha_{22}\end{pmatrix}\begin{pmatrix}\tau_1^{N+1}\\ \tau_2^{N+1}\end{pmatrix} = \begin{pmatrix}\beta_1 \\ \beta_2\end{pmatrix},
\end{equation}
where we have defined:
\begin{subequations}
\begin{align}
\alpha_{11} &\equiv\left[\triangle_A^{N+1}\Gamma_1^{N+1}-D_AT_{M-1}\right]\big|_{n=M-1},\\
\alpha_{12} &\equiv\left[\triangle_A^{N+1}\Gamma_2^{N+1}-D_AT_{M}\right]\big|_{n=M-1},\\
\alpha_{21} &\equiv\left[\triangle_A^{N+1}\Gamma_1^{N+1}-D_AT_{M-1}\right]\big|_{n=M},\\
\alpha_{22} &\equiv\left[\triangle_A^{N+1}\Gamma_2^{N+1}-D_AT_{M}\right]\big|_{n=M},\\
\beta_{1} &\equiv\left[\boldsymbol{\hat{\nabla}_A}^{N+1}\cdot\boldsymbol{\hat{v}}^{N+\frac{3}{4}}-\triangle_A^{N+1}\hat{\Pi}_0^{N+1}\right]\big|_{n=M-1},\\
\beta_{2} &\equiv\left[\boldsymbol{\hat{\nabla}_A}^{N+1}\cdot\boldsymbol{\hat{v}}^{N+\frac{3}{4}}-\triangle_A^{N+1}\hat{\Pi}_0^{N+1}\right]\big|_{n=M},
\end{align}
\end{subequations}
where the notation $\hat{q}\big|_{n=M}$ represents the $M$th Chebyshev mode
of function $\hat{q}$.
Note that if $M$ is even, then $\Gamma_1^{N+1}$ is an even polynomial,
$\Gamma_2^{N+1}$ is an odd polynomial; if $M$ is odd, then $\Gamma_1^{N+1}$
is odd and $\Gamma_2^{N+1}$ is even.  Using these symmetries, one can show
that $\alpha_{11}=\alpha_{22}=0$, and $\tau_1^{N+1} = \beta_2/\alpha_{21}$,
$\tau_2^{N+1}= \beta_1/\alpha_{12}$.

This method effectively triples the computational time for the enthalpy step
(three Poisson-like solves instead of one per timestep).  However, the
enthalpy step is still only of order 10\% of the total computational time,
as the FFTs in the nonlinear advection dominate the computation.  We also
note that if we were not working in Lagrangian coordinates (which introduced 
explicit time dependence into the gradient operators), then the Greens
functions $\Gamma_1$ and $\Gamma_2$ (two for each set of Fourier wavenumbers)
would be independent of time, and would only have to be computed once in a
pre-processing step, and stored.  

\section{Recursion relations for rational Chebyshev functions}

Here we present some useful recursion relations for the rational Chebyshev
functions.  We define:
\begin{align}
f(z) = \sum_{n=0}^{N}\hat{f}_n^C\: C_n(z) \quad\text{or}\quad f(z) = \sum_{n=0}^{N}\hat{f}_n^S\: S_n(z),\nonumber\\
g(z) = \sum_{n=0}^{N}\hat{g}_n^C\: C_n(z) \quad\text{or}\quad g(z) = \sum_{n=0}^{N}\hat{g}_n^S\: S_n(z).
\end{align}

The first derivative operator converts a $C_n(z)$ series into a $S_n(z)$
series, and vice versa.  The recursion relations have the same form
for either case, except for the lowest two modes.  Let $g=df/dz$, then:
\begin{align}
\hat{g}_n^{S/C} &= \left[-(n-2)\hat{f}_{n-2}^{C/S} + 2n\hat{f}_n^{C/S} - (n+2)\hat{f}_{n+2}^{C/S}\right]/(4L_z),\quad\text{for}\quad n\ge 2,\nonumber\\
\hat{g}_0^S &= 0, \quad \hat{g}_1^S = \left(3\hat{f}_1^C-3\hat{f}_3^C\right)/(4L_z),\quad \text{OR}\nonumber\\
\hat{g}_0^C &= \hat{f}_2^S/(2L_z),\quad \hat{g}_1^C = \left(-\hat{f}_1^S+3\hat{f}_3^S\right)/(4L_z).
\end{align}

The second derivative operator preserves the kind of expansion; that is, 
the second derivative of a $C_n(z)$ series is another $C_n(z)$ series,
and the second derivative of a $S_n(z)$ series is another $S_n(z)$ series.
The recursion relations for the second derivative operator have the same
form for either case, except for the lowest four modes.  Let $g=d^2f/dz^2$,
then:
\begin{align}
(16L_z^2)\hat{g}_n^{C/S} &= -(n-4)(n-2)\hat{f}_{n-4}^{C/S} + 4(n-2)(n-1)\hat{f}_{n-2}^{C/S} - 6n^2\hat{f}_n^{C/S} \nonumber \\
{}&+ 4(n+2)(n+1)\hat{f}_{n+2}^{C/S} - (n+4)(n+2)\hat{f}_{n+4}^{C/S}\quad\text{for}\quad n \ge 4,\nonumber\\
\hat{g}_0^C &= \left[8\hat{f}_2^C-8\hat{f}_4^C\right]/(16L_z^2),\nonumber\\
\hat{g}_1^C &= \left[-6\hat{f}_1^C+21\hat{f}_3^C-15\hat{f}_5^C\right]/(16L_z^2),\nonumber\\
\hat{g}_2^C &= \left[-24\hat{f}_2^C+48\hat{f}_4^C-24\hat{f}_6^C\right]/(16L_z^2),\nonumber\\
\hat{g}_3^C &= \left[9\hat{f}_1^C-54\hat{f}_3^C+80\hat{f}_5^C-35\hat{f}_7^C\right]/(16L_z^2),\quad\text{OR}\nonumber\\
\hat{g}_0^S &= 0,\nonumber\\
\hat{g}_1^S &= \left[-6\hat{f}_1^S+27\hat{f}_3^S-15\hat{f}_5^S\right]/(16L_z^2),\nonumber\\
\hat{g}_2^S &= \left[-24\hat{f}_2^S+48\hat{f}_4^S-24\hat{f}_6^S\right]/(16L_z^2),\nonumber\\
\hat{g}_3^S &= \left[7\hat{f}_1^S-54\hat{f}_3^S+80\hat{f}_5^S-35\hat{f}_7^S\right]/(16L_z^2).
\end{align}

Like the second derivative operator, the $z(d/dz)$ operator preserves the kind
of expansion.  Let $g=z(df/dz)$, then:
\begin{align}
\hat{g}_n^{C/S} &= \left[-(n-2)\hat{f}_{n-2}^{C/S}+(n+2)\hat{f}_{n+2}^{C/S}\right]/4 \quad\text{for}\quad n \ge 2,\nonumber\\
\hat{g}_0^C &= \hat{f}_2^C/2, \quad \hat{g}_1^C = \left[\hat{f}_1^C + 3\hat{f}_3^C\right]/4,\quad\text{OR}\nonumber\\
\hat{g}_0^S &= 0,\quad \hat{g}_1^S = \left[-\hat{f}_1^S + 3\hat{f}_3^S\right]/4,
\end{align}

One last important operation is multiplication by $z$.  A simple recursion
relation exists only for multiplication by $z$ on a $S_n(z)$ series, which
turns it into a $C_s(z)$ series.  Let $g(z)=zf(z)$, then:
\begin{equation}
\hat{g}_n^C = 
\begin{cases}
{\displaystyle\sum_{\substack{k=2\\\text{$k$ even}}}^{N}\hat{f}_k^S \quad\text{for $n=0$}},\\
{\displaystyle\hat{f}_n^S + 2\sum_{\substack{k=n\\\text{$k$ even}}}^{N}\hat{f}_k^S \quad\text{for $n$ even}},\\ 
{\displaystyle\hat{f}_n^S + 2\sum_{\substack{k=n\\\text{$k$ odd}}}^{N}\hat{f}_k^S \quad\text{for $n$ odd}}.\\ 
\end{cases}
\end{equation}

In practice, the recursion relations for $z(d/dz)$ and $d^2/dz^2$ are
unnecessary, as matrices for these operations can be built up from
the first derivative matrices and the multiplication by $z$ matrix.


\begin{thebibliography}{26}
\expandafter\ifx\csname natexlab\endcsname\relax\def\natexlab#1{#1}\fi
\expandafter\ifx\csname url\endcsname\relax
  \def\url#1{\texttt{#1}}\fi
\expandafter\ifx\csname urlprefix\endcsname\relax\def\urlprefix{URL }\fi

\bibitem[{Adams and Watkins(1995)}]{aw95}
Adams, F., Watkins, R., 1995. Vortices in circumstellar disks. Astrophys. J.
  451, 314--327.

\bibitem[{Bannon(1996)}]{bannon96b}
Bannon, P., 1996. On the anelastic approximation for a compressible atmosphere.
  J. Atmos. Sci. 53, 3618--3628.

\bibitem[{Barge and Sommeria(1995)}]{barge95}
Barge, P., Sommeria, J., 1995. Did planet formation begin inside persistent
  gaseous vortices{?} Astron. Astrophys. 295, L1--4.

\bibitem[{Barranco and Marcus(2000)}]{barranco00b}
Barranco, J., Marcus, P., 2000. Vorticies in protoplanetary disks and the
  formation of planetesimals. In: Center for {T}urbulence {R}esearch --
  {P}roceedings of the 2000 {S}ummer {P}rogram. pp. 97--108.

\bibitem[{Barranco and Marcus(2005)}]{barranco05a}
Barranco, J., Marcus, P., 2005. Three-dimensional vortices in stratified
  protoplanetary disks. Astrophys. J. 623, 1157--1170.

\bibitem[{Barranco et~al.(2000)Barranco, Marcus, and Umurhan}]{barranco00a}
Barranco, J., Marcus, P., Umurhan, M., 2000. Scalings and asymptotics of
  coherent vortices in protoplanetary disks. In: Center for {T}urbulence
  {R}esearch -- {P}roceedings of the 2000 {S}ummer {P}rogram. pp. 85--96.

\bibitem[{Boyd(1998)}]{boyd98}
Boyd, J., 1998. Two comments on filtering (artificial viscosity) for
  {C}hebyshev and {L}egendre spectral and spectral element methods: Preserving
  boundary conditions and interpretation of the filter as a diffusion. J. Comp.
  Phys. 143, 283--288.

\bibitem[{Boyd(2000)}]{boyd00}
Boyd, J., 2000. Chebyshev and Fourier Spectral Methods. Dover Publications,
  Inc., Mineola, NY.

\bibitem[{Cain et~al.(1984)Cain, Ferziger, and Reynolds}]{cain84}
Cain, A., Ferziger, J., Reynolds, W., 1984. Discrete orthogonal function
  expansions for non-uniform grids using the {F}ast {F}ourier {T}ransform. J.
  Comp. Phys. 56, 272--286.

\bibitem[{Canuto et~al.(1988)Canuto, Hussaini, Quarteroni, and Zang}]{canuto88}
Canuto, C., Hussaini, M., Quarteroni, A., Zang, T., 1988. Spectral Methods in
  Fluid Dynamics. Springer-Verlag, New York.

\bibitem[{Gilman and Glatzmaier(1981)}]{gilman81}
Gilman, P., Glatzmaier, G., 1981. Compressible convection in a rotating
  spherical shell. {I}. {A}nelastic equations. Astrophys. J., Suppl. Ser. 45,
  335--349.

\bibitem[{Goldreich and Lynden-Bell(1965)}]{goldreich65a}
Goldreich, P., Lynden-Bell, D., 1965. {I}. {G}ravitational stability of
  uniformly rotating disks. Mon. Not. R. Astron. Soc. 130, 97--124.

\bibitem[{Gottlieb and Orszag(1977)}]{gottlieborszag77}
Gottlieb, D., Orszag, S., 1977. Numerical Analysis of Spectral Methods: Theory
  and Applications. Society for Industrial and Applied Mathematics,
  Philadelphia.

\bibitem[{Gough(1969)}]{gough69}
Gough, D., 1969. The anelastic approximation for thermal convection. J. Atmos.
  Sci. 26, 448--456.

\bibitem[{Kida(1981)}]{kida81}
Kida, S., 1981. Motion of an elliptic vortex in a uniform shear flow. J. Phys.
  Soc. Jpn. 50~(10), 3517--3520.

\bibitem[{Kundu(1990)}]{kundu90}
Kundu, P., 1990. Fluid Mechanics. Academic Press, Inc., San Diego.

\bibitem[{Marcus(1984)}]{marcus84a}
Marcus, P., 1984. Simulation of {T}aylor--{C}ouette flow. {I}. {N}umerical
  methods and comparison with experiment. J. Fluid Mech. 146, 45--64.

\bibitem[{Marcus(1986)}]{marcus86a}
Marcus, P., 1986. Description and philosophy of spectral methods. In: Winkler,
  K.-H., Norman, M. (Eds.), Proceedings of Astrophysical Radiation
  Hydrodynamics. Springer-Verlag, pp. 359--386.

\bibitem[{Marcus(1990)}]{marcus90a}
Marcus, P., 1990. Vortex dynamics in a shearing zonal flow. J. Fluid Mech. 215,
  393--430.

\bibitem[{Marcus(1993)}]{marcus93}
Marcus, P., 1993. Jupiter's {G}reat {R}ed {S}pot and other vortices. Annu. Rev.
  Astron. Astrophys. 31, 523--573.

\bibitem[{Marcus and Press(1977)}]{marcus77}
Marcus, P., Press, W., 1977. On {G}reen's functions for small disturbances of
  plane {C}ouette flow. J. Fluid Mech. 79, 525--534.

\bibitem[{Miesch et~al.(2000)Miesch, Elliott, Toomre, Clune, Glatzmaier, and
  Gilman}]{miesch00}
Miesch, M., Elliott, J., Toomre, J., Clune, T., Glatzmaier, G., Gilman, P.,
  2000. Three-dimensional spherical simulations of solar convection. {I}.
  {D}ifferential rotation and pattern evolution achieved with laminar and
  turbulent states. Astrophys. J. 532, 593--615.

\bibitem[{Moore and Saffman(1971)}]{ms71}
Moore, D., Saffman, P., 1971. Structure of a line vortex in an imposed strain.
  In: Olsen, J.~H., Goldburg, A., Rogers, H. (Eds.), Aircraft Wake Turbulence
  and its Detection. Plenum Press, New York, pp. 339--354.

\bibitem[{Ogura and Phillips(1962)}]{ogura62}
Ogura, Y., Phillips, N., 1962. Scale analysis of deep and shallow convection in
  the atmosphere. J. Atmos. Sci. 19, 73--79.

\bibitem[{Pedlosky(1979)}]{pedlosky79}
Pedlosky, J., 1979. Geophysical Fluid Dynamics. Springer-Verlag, New York.

\bibitem[{Rogallo(1981)}]{rogallo81}
Rogallo, R., 1981. Numerical experiments in homogeneous turbulence. Technical
  memorandum 81315, NASA.

\end{thebibliography}

\end{document}